\documentclass[fleqn,usenatbib]{mnras}

\usepackage{newtxtext,newtxmath}
 
\usepackage[T1]{fontenc}

\DeclareRobustCommand{\VAN}[3]{#2}
\let\VANthebibliography\thebibliography
\def\thebibliography{\DeclareRobustCommand{\VAN}[3]{##3}\VANthebibliography}

\usepackage{graphicx}	
\usepackage{amsmath}	
\usepackage{amssymb}	
\usepackage[labelfont=bf]{caption}

\newcommand{\kmsmpc}{\kms\;{\rm Mpc}^{-1}}

\newcommand{\hkpc}{h^{-1}{\rm kpc}}
\newcommand{\hmpc}{h^{-1}{\rm Mpc}}

\newcommand{\kms}{\;{\rm km}\,{\rm s}^{-1}}

\newcommand{\gizmo}{{\sc Gizmo}}

\newcommand{\simba}{{\sc Simba}}

\newcommand{\fgas}{f_{\rm gas}}
\newcommand{\fedd}{f_{\rm Edd}}
\newcommand{\mbh}{\;{M}_{\rm BH}}
\newcommand{\mstar}{\;{M}_{*}}

\newcommand{\ro}{R_{0}}
\newcommand{\msun}{\;{\rm M}_{\odot}}
\newcommand{\mhalo}{\;{M}_{\rm halo}}
\newcommand{\mdot}{\;{\dot{M}}_{\rm BH}}

\newcommand{\WHz}{\rm{~W~Hz}^{-1}}
\newcommand{\power}{\;P_{\rm 1.4~GHz}}
\newcommand{\pmhz}{\;P_{\rm 150~MHz}}
\newcommand{\edit}{\textcolor{black}}

\title[Environments of radio galaxies in \simba]{The environments of the radio galaxy population in {\sc SIMBA}}

\author[Thomas \& Dav\'{e}]{
Nicole Thomas,$^{1,2,3}$\thanks{E-mail: thomas.nicolelynn@gmail.com}
Romeel Dav\'{e},$^{3,4,5}$
\\
$^{1}$Institute for Computational Cosmology, Department of Physics, University of Durham, South Road, Durham DH1 3LE, UK\\
$^{2}$Centre for Extragalactic Astronomy, Department of Physics, Durham University, Durham DH1 3LE, UK\\
$^{3}$Department of Physics and Astronomy, University of the Western Cape, Bellville, 7535, South Africa\\
$^{4}$Institute for Astronomy, Royal Observatory, University of Edinburgh, Edinburgh, EH9 3HK, UK\\
$^{5}$South African Astronomical Observatories, Observatory, Cape Town 7925, South Africa\\
}

\date{Accepted XXX. Received YYY; in original form ZZZ}

\pubyear{2020}

\begin{document}
\label{firstpage}
\pagerange{\pageref{firstpage}--\pageref{lastpage}}
\maketitle

\begin{abstract}
We investigate the environmental properties of the $z=0$ radio galaxy population using the \simba\ cosmological hydrodynamic simulation. We identify centrals and satellites from a population of high and low excitation radio galaxies (HERGs and LERGs) in \simba, and study their global properties. We find that $\sim$20$\%$ of radio galaxies are satellites, and that there are insignificant differences in the global properties of LERGs based on their central/satellite classification. HERG satellites display lower values of star formation, 1.4\,GHz radio luminosity, and Eddington fractions than HERG centrals. We further investigate the environments of radio galaxies and show that HERGs typically live in less dense environments, similar to star-forming galaxies. The environments of high-mass LERGs are similar to non-radio galaxies, but low-mass LERGs live in underdense environments similar to HERGs. LERGs with over-massive black holes reside in the most dense environments, while HERGs with over-massive black holes reside in underdense environments. The richness of a LERG’s environment decreases with increasing Eddington fraction, and the environments of all radio galaxies do not depend on radio luminosity for $\power < 10^{24} \WHz$. Complementing these results, we find that LERGs cluster on the same scale as the total galaxy population, while multiple HERGs are \edit{not} found within the same dark matter halo. Finally, we show that high density environments support the growth of HERGs rather than LERGs at $z=2$. \simba\ predicts that with more sensitive surveys, we will find populations of radio galaxies in environments much similar to the total galaxy population.
\end{abstract}




\section{Introduction}
It has long been known that the supermassive black holes~(SMBHs) at the centres of most massive galaxies co-evolve with their hosts~\citep{FerrareseMerritt2000,Gebhardt2000,KH2013}. Actively accreting SMBHs, or active galactic nuclei~(AGN), can be classified in many ways, with one particularly useful physically motivated classification being that into ``radiative mode'' and ``jet mode''~\citep{HeckmanAndBest2014} AGN. 

Radiative mode (alternatively, ``cold mode'', ``quasar mode'', or ``high excitation mode'') AGN are typically observed to have $\fedd\ga 0.01$, where the Eddington fraction $\fedd$ is the accretion rate divided by the Eddington rate for given black hole of mass ($\mbh$), and accrete via a geometrically thin, optically thick disk of cold gas~\citep{Shakura1973}. These AGN are often surrounded by a dusty obscuring structure and emit radiation efficiently across the electromagnetic spectrum. Jet mode (alternatively, ``hot mode'', ``radio mode'', or ``low excitation mode'') AGN typically have an inefficient advection dominated accretion flow~(ADAF) via a spherical hot medium~\citep{Hardcastle2007} with $\fedd\la 0.01$, and emit predominantly at radio wavelengths in the form of mechanical jets. While radiative mode AGN are typically dominated by emission from the accretion disk, a small fraction of radiative mode AGN are observed to produce powerful radio jets. These AGN are referred to as High Excitation Radio Galaxies (HERGs) and are classified by the presence of high excitation lines in their optical spectra~\citep{BestandHeckman2012,Whittam2016,Whittam2018}. The inefficiently-accreting jet mode AGN are often classified as Low Excitation Radio Galaxies (LERGs) owing to their lack of high excitation optical lines.

Radio galaxies with AGN-driven jets have traditionally been associated with galaxy clusters~\citep{PrestagePeacock1988} and, particularly in the nearby universe, prefer rich environments~\citep{Best2004}. Galaxy groups and clusters have a high probability of hosting a radio galaxy as their central or brightest cluster galaxy~(BCG)~\citep{Best2007}.  A caveat, however, is that this might owe primarily to the large size of the BCG (and hence its black hole) rather than the intrinsic nature of the radio galaxy~\citep{Best2005b}. \citet{Best2007} shows that, while BCGs may be more likely to host a radio galaxy, a radio galaxy can also be hosted by other cluster galaxies; despite this, brightest group or cluster galaxies are \textit{more likely} to host a radio loud AGN than other group or cluster galaxies spanning the same stellar mass~\citep{Moravec2020,GoldenMarx2021}.

The relationship between radio galaxies and their host clusters can also be probed at X-ray wavelengths, via X-ray emission tracing the hot intra-cluster medium (ICM). Using X-ray observations with {\it Chandra} and {\it XMM-Newton}, \citet{Ineson2015} shows that there are strong correlations between the radio luminosity and cluster X-ray luminosity for LERGs, though not for HERGs; however, these samples are restricted to the most X-ray luminous clusters. It is expected that the radio luminosity of the central radio galaxy scales with the X-ray luminosity of the cluster~\citep{Pasini2020,Pasini2021}, implying that brighter radio galaxies populate richer environments~\citep{Best2004,Ching2017,Croston2018}. 

The cooling rates in more massive clusters are lower, thus the accretion of hot gas from the ICM onto the black hole is expected to be less efficient. This implies that there will be reduced amounts of cold gas accreted onto the SMBH at the center of the central radio galaxy in massive halos, suggesting that richer environments can be used as an indicator of the mode of accretion of the central radio galaxy~\citep{Ineson2015,Croston2018}. 
In support of this,~\citet{Hale2018} showed that low Eddington rate AGN are more clustered than high Eddington rate AGN, showing that more massive clusters reduce the efficiency of accretion in AGN and additionally suppress star formation. They further find that AGN cluster more strongly than star forming galaxies (SFGs). If the high and low accretion efficiency populations are traced by HERGs and LERGs, they would be expected to live in different environments, with those at higher radio luminosities living in more dense environments. Indeed it is found that high luminosity LERGs live in denser environments, while the environments of HERGs and low luminosity LERGs are indistinguishable ~\citep{Ching2017}.

The variation in the morphological classifications of radio galaxies are also thought to be due to the accretion mode of the central SMBH, and thus to environment. \citet{Croston2019} showed that for classes of Fanaroff and Riley~(FR) radio galaxies, edge-darkened FRI's inhabit richer environments than edge-brightened FRII's. They also showed that only a small fraction of radio loud AGN at low redshift have a high probability of being associated to clusters, which implies that majority of low redshift AGN inhabit dark matter haloes with $\mhalo < 10^{14} \msun$. In addition they find that the probability of being associated with a cluster relates closely to AGN radio luminosity and is not driven only by the galaxy stellar mass. This result supports the idea that large scale environment is a significant role player in the driving of jet activity in low redshift AGN.

On the theoretical front, models suggest that radio jets are a key player in enacting and/or maintaining the lack of star formation in massive galaxies~\citep{Bower2006,Croton2006} that tend to live in dense environments. In galaxy clusters, radio jets provide enough energy into the ICM to suppress the cooling of hot gas, thereby giving rise to a population of quenched galaxies~\citep{Fabian1999,Fabian2003,Best2005b,McNamara2007}. Moreover, the hot gas from the ICM is what is thought to fuel the accretion onto the SMBH at the center of the radio galaxy, and thus radio galaxies may be connected to the environment because they tend to be ``jet mode'' AGN.

Therefore, studying the environments of radio galaxies is important in understanding how the activity of SMBHs are started and fuelled over their lifetime, and the impact the environment might play in their evolution. Understanding this better could, for instance, help use radio galaxies as locators of high density environments tracing large scale structure.

Cosmological simulations have been beneficial in understanding the connection between galaxies and their surrounding dark matter halos, and have been further successful in describing the properties of galaxies as a function of their environments. This owes in large part to the inclusion of SMBH growth and AGN feedback, which quenches massive galaxies in dense regions~\citep{Somerville2015}. Most modern simulations grow their SMBHs solely via Bondi accretion~\citep{Bondi1952,BondiAndHoyle,Hoyle1939} which describes the spherical gravitational accretion from a hot medium around the black hole, and thus better models the growth of SMBHs in massive quenched systems with little cold gas such as galaxies in clusters and LERGs.

Simulations have widely differing prescriptions for AGN feedback. Using a model with thermal feedback in which energy is stochastically deposited thermally and isotropically around the black hole, the C-EAGLE simulation~\citep{Barnes2017} successfully reproduced the stellar and black hole mass properties of clusters as well as their metal content and Sunyaev-Zeldovich properties. However, the clusters are too gas rich, implying that the feedback coming from AGN at earlier times is not strong enough to drive out gas from the cluster progenitor, while simultaneously the central temperature of the clusters are too high indicating that the AGN feedback is too strong once the core of the cluster has been formed.

Using the IllustrisTNG simulations that include a kinetic jet model ejected bipolarly in a random direction at each time-step, \citet{Bhowmick2020} showed that AGN activity is enhanced during mergers, but that this only accounts for $\sim$11$\%$ of the AGN population. \citet{Truong2021} found strong correlations between the X-ray luminosity of clusters with the mass of SMBHs in quenched galaxies, and showed that this relation is sensitive to the nature of ongoing feedback processes and thus dependent on the availability of gas accreted on the SMBH. Along those lines,~\citet{Barnes2019} shows that including anisotropic thermal conduction into the ICM alongside AGN feedback may be an important process to reproduce the thermal structure of the ICM.  These results highlight the interplay of AGN feedback and environment probed in models.

The \simba\ suite of simulations differs from most others in both its accretion and feedback models. \simba\ is unique among recent cosmological simulations in that it includes the torque-limited accretion model~\citep{DAA2013,DAA2015} when accreting cooler gas, and only uses Bondi accretion for hot gas. For AGN feedback, it includes stably bipolar kinetic radiative and jet modes distinguished via the SMBH's Eddington rate, in a manner consistent with the observationally-motivated scenario in~\citet{HeckmanAndBest2014}. This yields a galaxy population in accord with a wide range of observations~\citep[e.g.][]{Dave2019} as well as galaxy--SMBH co-evolution~\citep{Thomas2019,Habouzit2020}. In addition, the X-ray halo properties in \simba\ match well to observations including the baryon fractions as a function of halo mass, and indicate that AGN jet feedback is responsible for the evacuation of hot gas within group-sized halos~\citep{Robson2020,Appleby2021}.  Thus \simba\ provides an interesting platform to examine environmental effects on the AGN population.

A nice benefit of \simba\ is that its jet mode feedback can naturally be associated with radio galaxies, while its two accretion modes naturally map onto high and low excitation radio galaxies (HERGs and LERGs).  The unique subgrid models thus allow \simba\ to be the first simulation to classify simulated AGN into HERGs and LERGs in a physically motivated way. \citet{Thomas2021} showed that the bulk properties of HERGs and LERGs predicted in this way are a good match to observations at $z=0$ and earlier epochs.  The environmental dependence of radio galaxies offers another way to test and constrain these models, as well as gain insights into the connection between environment, black hole growth, and galaxy quenching.

In this paper we use \simba\ to investigate the environments of radio galaxies, in particular we investigate the differences in properties of central and satellite radio galaxies as well as study the clustering properties of radio galaxies. This work follows on from the separation of radio galaxies into HERGs and LERGs in~\citet{Thomas2021} and aims to understand the role environments have on the evolution of radio galaxies. 

This paper is prepared as follows. In \S\ref{sec:sims} we discuss the \simba\ simulations and the black hole accretion (\S\ref{sec:accretion}) and feedback (\S\ref{sec:feedback}) models implemented therein. In \S\ref{sec:cvs} we compare the properties of central and satellite radio galaxies and in \S\ref{sec:env} we investigate the clustering properties of radio galaxies and how this differs to $z=0$ at $z=2$. We summarise our findings in \S\ref{sec:conclude}.

\section{Simulations}
\label{sec:sims}
\simba~\citep{Dave2019} is a state-of-the-art suite of cosmological hydrodynamic simulations which uses a branched version of the \gizmo\ cosmological gravity and hydrodynamics solver~\citep{Hopkins2015} in its Meshless Finite Mass (MFM) mode. \simba\ is particularly unique in its prescription for black hole growth in that it accounts for the accretion from both hot gas via Bondi accretion as well as cold gas via gravitational torque limited accretion. Additionally, \simba\ accounts for feedback from black holes via kinetic winds, jets, and high energy X-rays. 
A full description of \simba\ can be found in~\citet{Dave2019} along with more fundamental simulation properties listed in Table 1 therein, and a more detailed description of the radio galaxy population can be found in~\citet{Thomas2021}.
Due to the relevance of radio galaxies in this work, below we recap the modelling of black holes within \simba, as well as the theoretical models used to estimate the 1.4\,GHz radio luminosities of radio galaxies in \simba. 

We use \simba's fiducial $(100 \hmpc)^{3}$ box containing $1024^{3}$ dark matter particles and $1024^{3}$ gas elements evolved from $z= 249 \to 0$ assuming a \citet{Planck2016} concordant cosmology with $\Omega_{m}=0.3$, $\Omega_{\Lambda}=0.7$, $\Omega_{b}= 0.048$, $H_{0}=0.68\kmsmpc$, $\sigma_{8}=0.82$, and $n_{s}=0.97 $. The \simba\ mass resolution for dark matter and gas is $m_{\rm DM} = 9.6\times 10^{7} \msun$ and $m_{\rm gas} = 1.82 \times 10^{7} \msun$ respectively. 

\simba\ identifies halos on the fly during the simulation via a friends-of-friends algorithm with a linking length 0.2 times the mean inter-particle spacing. This is applied to all stars, black holes, and gas elements with a density above that of the star formation density threshold of $n_{H}>0.13$ H atoms~cm$^{-3}$. Above the star formation density threshold, \simba\ stochastically spawns star particles from gas elements with the same mass. Galaxies are resolved down to 32 star particles, above which galaxies properties are computed reliably. This resolution is equivalent to a minimum galaxy stellar mass of $\mstar=5.8\times10^{8}\msun$. 

Galaxies at a stellar mass of $\mstar \geq \gamma_{\rm BH} \mbh$ with $\gamma_{\rm BH} = 3 \times 10^{5}$ are seeded with black holes of $\mbh = 10^{4} h^{-1}\msun $, based on the idea that small galaxies suppress black hole growth via stellar feedback~\citep{Dubois2015,Bower2017,DAA2017b,Habouzit2017,Hopkins2021}. 
This seeding of black holes occur then for galaxies with $\mstar \approx 10^{9.5} \msun$, i.e. $\approx 175$ star particles. In this work we will generally be considering more massive galaxies that host jet mode AGN feedback.

The publicly available {\sc yt}-based package {\sc Caesar}\footnote{\tt https://caesar.readthedocs.io}~\citep{YT,Thompson_ASCL} is used to compute properties of galaxies which include black hole and dark matter halo properties of the galaxies, outputting a catalogue in HDF5 format for each simulation snapshot. In \simba\ we define the central galaxy in a dark matter halo as the highest stellar mass galaxy in that halo. All other galaxies within the same halo are identified as satellites.

This work will primarily be focused on $z=0$ with a brief consideration at $z=2$. Next we briefly detail the sub-resolution models implemented to describe black hole accretion and feedback within \simba.

\subsection{Black Hole Accretion}
\label{sec:accretion}

\simba\ employs a two-mode accretion model for black hole growth: the gravitational torque-limited accretion model~\citep{DAA2017} from cold gas ($T<10^{5}K$), and Bondi accretion~\citep{BondiAndHoyle,Bondi1952,Hoyle1939} from hot gas ($T>10^{5}K$).

\subsubsection{Gravitational Torque-Limited Model}
\label{sec:gt}
Accretion rates for cold gas are based on the gravitational torque model of \citet{HopkinsQuataert2011} which estimates the gas inflow rate, $\dot{\rm M}_{\rm Torque}$, driven by gravitational instabilities from galactic scales down to the accretion disk surrounding the black hole. This model scales as $\mbh^{1/6}$ and depends on the gas, disk, and stellar masses within a distance $\ro$ of the black hole. $\ro$ encloses 256 gas elements with an upper limit of $2\hkpc$. 
Crucially, \citet{HopkinsQuataert2011} demonstrated using parsec-scale isolated disk simulations that this model yields consistent results even for $\ro$ as large as a kpc, which is comparable to the resolution in \simba.  This suggests that although far from able to resolve the relevant disk instabilities, \simba\ will still broadly capture the behaviour of torque-limited black hole accretion.

\subsubsection{Bondi-Hoyle-Lyttleton Parameterisation}
\label{sec:bondi}

The Bondi model is a prescription for black hole growth widely used in galaxy formation simulations \citep[e.g.][]{Springel2005,Dubois2012,Choi2012}. The Bondi model describes the accretion of a black hole with mass $\mbh$ moving at a velocity $v$ relative to a uniform distribution of gas with density $\rho$ and sound speed $c_{s}$ and scales as $\mbh^{2}$. For comparison to a real black hole, the Milky Way's black hole is accreting at around the Bondi rate, but the advection-dominated accretion flow (ADAF) rate is somewhat less~\citep{Quataert1999}. In \simba, we aim to be modeling this ADAF rate via Bondi accretion from hot gas. 

\subsubsection{Numerical Implementation}
For all gas within $\ro$ that has a temperature $T<10^5$K, we apply the torque-limited accretion prescription, while for $T>10^5$K gas we employ the Bondi prescription computing $\rho$ and $c_{s}$ from hot gas only within $\ro$. Therefore, a given black hole can accrete gas via both Bondi and torque-limited modes at any given timestep. The total accretion onto the black hole is then
\begin{equation}
\label{eq:totaccr}
\dot{M}_{\rm BH}=(1-\eta)(\dot{M}_{\rm Bondi}+\dot{M}_{\rm Torque})
\end{equation}
where $\eta=0.1$ is the radiative efficiency of the black hole. Bondi accretion is limited to the Eddington rate, while torque-limited accretion is capped at $3\times$ the Eddington rate due to its lack of spherical symmetry. 

Black hole accretion proceeds stochastically~\citep{Springel2005}. Gas particles within $\ro$ get a fraction of their mass subtracted and added to the black hole, with a probability that statistically satisfies the mass growth (equation \ref{eq:totaccr}).  If a particle is sufficiently small compared to its original mass, it is swallowed completely.  Black holes are merged if they come within each other's $R_0$, since it is not possible to resolve the dynamical friction processes that lead to merging.

\subsection{Black Hole Feedback}
\label{sec:feedback}

Motivated by the observed dichotomy of accretion rates of AGN~\citep{HeckmanAndBest2014} and their corresponding outflows, \simba\ employs a multi-mode feedback model governed by the instantaneous Eddington ratio of the black hole. 
\subsubsection{Kinetic Feedback}
There is significant evidence suggesting that radiatively efficient AGN can drive strong winds~\citep{Sturm2011,Fabian2012}. For high $\fedd$ mode outflows, the outflow velocity of these winds can be estimated by the ionised gas linewidths of X-ray detected AGN from SDSS observations~\citep{Perna2017a}.
We refer to these outflows as AGN winds.

If $\fedd<0.2$, we slowly begin transitioning to the jet mode where the velocity becomes increasingly higher as $\fedd$ drops.
The velocity increase is capped at 7000$\kms$ and results in a maximum jet speed at $\fedd\leq0.02$. An additional criterion requiring $\mbh>{\rm M}_{\rm BH,lim}$ is added and motivated by observations that show that jets only arise in galaxies with velocity dispersions corresponding to black holes with $\mbh\gtrsim 10^{8} \msun$~\citep{Barisic2017,Mclure2004}; we conservatively choose ${\rm M}_{\rm BH,lim}=10^{7.5} \msun$. 

AGN-driven outflows are modelled by stochastically kicking particles near the black holes with velocity $v_w$ with a probability based on the kernel weight and  fraction of mass accreted by the black hole and subtracted from the gas particle before ejection~\citep{DAA2017}. 
The momentum outflow choice is based on the inferred energy and momentum inputs from observations of AGN outflows~\citep{Fiore2017,Ishibashi2018}, albeit towards the upper end of the observations, which we found is required in \simba\ in order to enact sufficient feedback in \simba\ to quench galaxies.


\subsubsection{X-ray Feedback}

\simba\ additionally includes high-energy photon pressure feedback, although this does not play a significant role in the present work.  The energy input rate due to X-rays emitted by the accretion disk is computed following \citet{Choi2012}, assuming a radiative efficiency $\eta=0.1$.  In gas-rich galaxies, severe radiative losses are expected in the ISM, hence we only apply X-ray feedback below a galaxy gas fraction threshold of $\fgas<0.2$, and in galaxies with full velocity jets ($v_{\rm w}\ga 7000\kms$). The feedback is applied spherically within $\ro$, providing an outwards kick for star-forming gas and heating for non-starforming gas. This mode is important in understanding the hot gas in galaxy groups and clusters~\citep{Robson2020} and for providing a final evacuation of gas to fully quench galaxies~\citep{Dave2019}, which for instance manifests in green valley galaxy profiles and qualitatively improves agreement with observations~\citep{Appleby2019}.

\subsection{Modeling Radio Emission}
\label{sec:radio_em}

\subsubsection{1.4~GHz radio power}

We compute the 1.4~GHz radio emission $\power$ for all resolved galaxies in \simba\ considering radio emission both due to star formation and accretion by the central SMBH.
Within the AGN population, we define HERGs and LERGs as those SMBHs with accretion rates dominated by gravitational torque limited accretion and Bondi accretion respectively.  We briefly describe these aspects in more detail below; for a full description and motivation see \citet{Thomas2021}.

To compute $\power$, we account for the contributions from both star formation (mainly owing to synchrotron emission from supernova shocks), as well as black hole accretion. For star formation, we follow \citet{Condon1992} and tie it to the star formation rate from a \citet{Chabrier2003} initial mass function (IMF) for stars more massive than $5 \msun$, by summing the thermal and non-thermal contributions via 
\begin{equation}
    \frac{P_{\rm non-thermal}}{\rm W\ Hz^{-1}} = 5.3\times 10^{21} \left(\frac{\nu}{\rm GHz}\right)^{-0.8} \left(\frac{SFR_{\geq 5\msun}}{\msun {\rm yr^{-1}}}\right)
\end{equation}
\begin{equation}
    \frac{P_{\rm thermal}}{\rm W\ Hz^{-1}} = 5.5\times 10^{20} \left(\frac{\nu}{\rm GHz}\right)^{-0.1} \left(\frac{SFR_{\geq 5\msun}}{\msun \rm yr^{-1}}\right)
\end{equation}
where $\nu=1.4$~Ghz is the observed frequency, and SFR$_{\geq5\msun}$ is the star formation rate in stars more massive than $5\msun$. 

The AGN $\power$ luminosity comes from radio galaxies identified in \simba, where we define a radio galaxy as one that has ongoing jet feedback from its SMBH.  For this to happen, the galaxy must contain a black hole, which we ensure by the criterion $\mstar \geq 10^{9.5}\msun$ (where we seed BHs), and must be emitting a jet, which we set by the criteria $\mbh>10^{8} \msun$ and $0<\fedd\leq 0.02$ (where \simba\ jets reach full velocity).

To estimate $\power$ from these AGN, we sum the contributions from core and extended emission by applying empirical relations  that connect black hole accretion rate to the 1.4~GHz radio luminosity.  We note that, at the luminosities and number densities probed within the \simba\ volume, the vast majority of AGN are expected to be core-dominated; the space density of lobe-dominated systems such as Fanaroff-Riley Type II (FRII) objects is such that we would expect less than 1 in our entire volume.  Nonetheless we will account for a lobe contribution in some objects, though the results are largely insensitive to this.

For core emission, we follow \citet{Slyz2015} and employ the following empirical relations from \citet{Kording2008}:
\begin{equation}
    \label{eq:rad_all}
    \frac{P_{\rm Rad}}{10^{30}\rm erg\ s^{-1}} = \left( \frac{\mdot }{4\times10^{17}\rm g\ s^{-1}} \right)^{\frac{17}{12}},
\end{equation}
where ${P}_{\rm Rad}\sim \nu { P}_{\nu}$ where $\nu$ is the frequency of the radio emission, in this case $\nu=1.4$\,GHz.  In detail, the core emission likely comes from a region $\sim$tens of pc around the SMBH, which is unresolvable in \simba; we assume the contribution to $\power$ from the remainder of the galaxy (other than from star formation) is small in comparison.

The lobe contribution is more ad hoc, since we do not yet have the ability to track and identify such lobes in \simba. We select 10\% of HERGs that are central galaxies and that reside in the largest dark matter halos, and add a contribution given by:
\begin{equation}
    \label{eq:rad_hergs}
    \log\ P_{\rm 151} \left({\rm W\ Hz^{-1} sr^{-1}}\right) = \log\ \mdot  +0.15
\end{equation}
using ${P}_{\nu}\propto \nu^{-\alpha}$ to scale from $P_{\rm151~MHz}$ to $\power$ using the accepted spectral index of $\alpha=0.7$ for synchrotron emission. We note that this additional contribution does not make a significant impact to our results as this only accounts for 11/113 HERGs or 11/1365 radio galaxies. The effect of this contribution can be seen in \citet[Appendix A]{Thomas2021}. In general the AGN contribution dominates the radio luminosity for radio galaxies, however a fraction of the radio galaxies with high star formation rates are dominated by the respective star formation contribution.

\subsubsection{HERGs and LERGs}

We separate the radio galaxy population into high and low excitation radio galaxies (HERGs and LERGs) based on their dominant mode of accretion:  HERGs are assumed to occur in SMBH dominated by gravitational torque limited accretion (i.e. $>50\%$ contribution), while LERGs are those SMBH dominated by Bondi accretion. To account for the stochasticity of the accretion model in \simba\ we compute the average accretion rate over 50~Myr.  In \citet{Thomas2021} we show that the large majority of galaxies are strongly dominated by one or the other accretion mode with few ``in between" galaxies, so the results are insensitive to the exact threshold chosen to separate HERGs and LERGs.  By these prescriptions, \simba\ produces 1252 LERGs and 113 HERGs at $z=0$\footnote{We note a difference in number of HERGs and LERGs galaxies compared to \citet{Thomas2021}. This is merely due to an improvement in the 50Myr timescale averaging and makes no other significant change to the results provided.}. The ratio of number of HERGs to number of LERGs is consistent with observations in that radio galaxies, as expected from their high masses, are often hot-mode dominated.

We note that we do not explicitly tie the accretion mode to the feedback mode or to radio emission.  So for instance there is nothing that forces Bondi-dominated SMBH (LERGs) to be associated with strong jets or quenched galaxies, or HERGs to be in star-forming systems.  Nonetheless, such trends do emerge naturally in \simba\ owing to the interplay of jets, hot gas, and star formation~\citep{Thomas2021}.

\subsubsection{Global Properties of Radio Galaxies in \simba}
In \citet{Thomas2021} we define a population of radio galaxies based on the above criteria. Briefly, we found that, while radio galaxies are typically hosted by more massive galaxies in \simba, HERGs typically reside in mildly lower $\mstar$, $\mbh$, $\sigma_{*}$, $\mhalo$ and higher $sSFR$ galaxies relative to the most massive LERGs. HERGs span the full range of $sSFR$s while LERGs span a range of quenched to green valley galaxies. We additionally see a vast overlap in $\fedd$ of radio galaxies, which is in disagreement with a canonically observed dichotomy but agrees somewhat better with more recent observations showing more overlap in this property. 
We predict from these previous works that upcoming surveys will probe the fainter population of these sources and uncover increasingly less difference between the global properties HERGs and LERGs.  In this work, we employ this same population of radio galaxies identified and quantified in \citet{Thomas2021}.

\begin{figure}
    \centering
    \includegraphics[width=\linewidth]{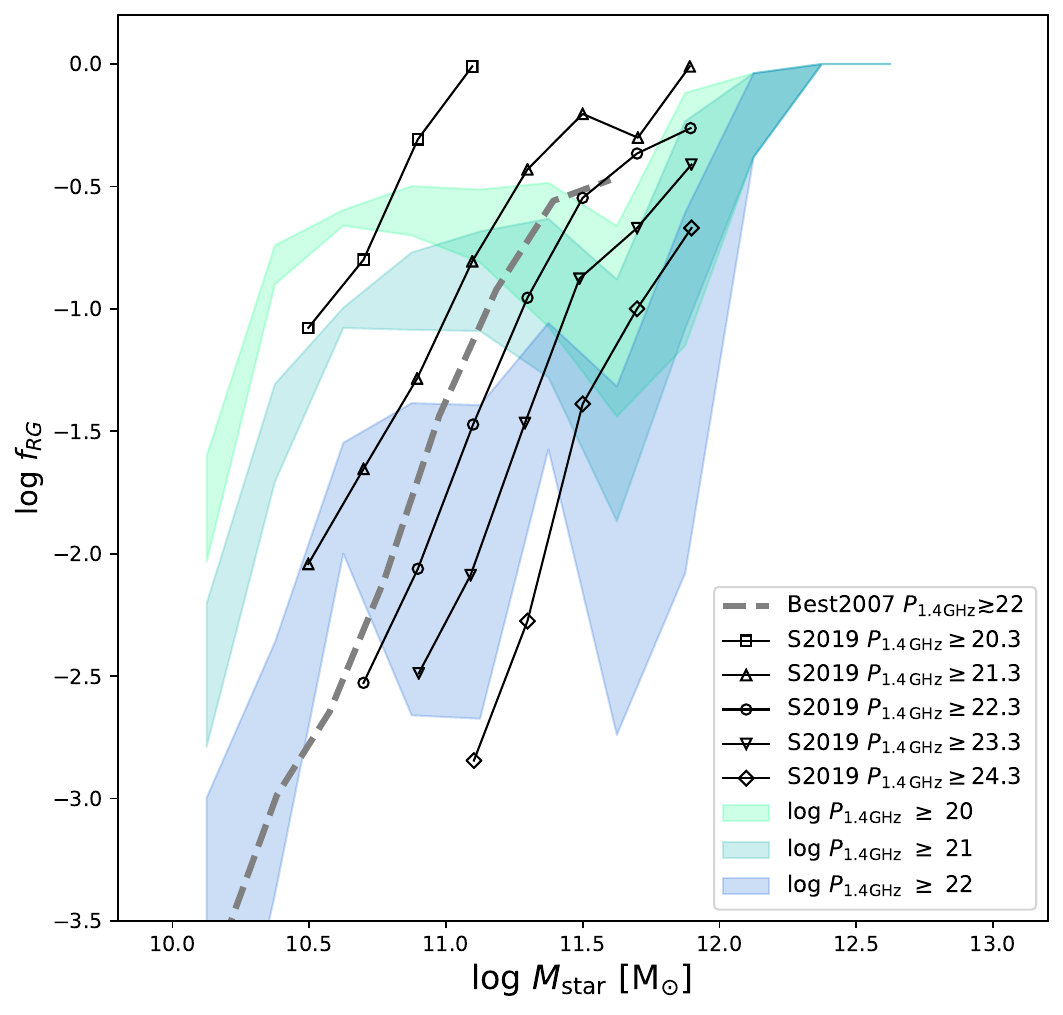}
    \captionof{figure}{The 1$\sigma$ error over 8 simulations suboctants of the fraction of radio galaxies in \simba\ as a function of stellar mass split into 3 luminosity bins; $\power \geq 10^{20} \WHz$ (green), $\power \geq 10^{21} \WHz$ (green-blue),and $\power \geq 10^{22} \WHz$ (blue), in comparison with observations of $\power \geq 10^{22} \WHz$ by \citet{Best2007} (grey dashed line) as well as $\power\geq 10^{20.3} \WHz$, $\power\geq 10^{21.3} \WHz$, $\power\geq 10^{22.3} \WHz$, $\power\geq 10^{23.3} \WHz$, $\power\geq 10^{24.3} \WHz$ (solid lines with empty squares, triangles, circles, downward triangles, and diamonds respectively) by \citet{Sabater2019}. \simba\ broadly reproduces the observed fraction of radio galaxies but underpredicts fractions between $10^{11} \leq \mstar [\msun] \leq 10^{11.5}$.}
    \label{fig:fagn}
\end{figure}

\subsubsection{Radio fraction vs. stellar mass}

Figure~\ref{fig:fagn} shows the the fraction of radio galaxies as a function of stellar mass in \simba, with bands indicating the $\pm$1$\sigma$ spread over 8 simulations sub-octants as a measure of cosmic variance. The population of radio galaxies are split into three luminosity bins with $\power \geq 10^{20} \WHz$ (green), $\geq 10^{21} \WHz$ (green-blue), and $\geq 10^{22} \WHz$ (blue). This is compared with observations by \citet{Best2007} for radio galaxies with $\power \ga 10^{22} \WHz$ as well as with \citet{Sabater2019} for radio galaxies with $\pmhz \ga 10^{21} \WHz$, $\ga 10^{22} \WHz$, $\ga 10^{23} \WHz$, $\ga 10^{24} \WHz$, and $\ga 10^{25} \WHz$ (lines with various point types). Note that we do not expect \simba\ to match these observations precisely as they correspond to 1.4\,GHz radio luminosities of corresponding to $\power \ga 10^{20.3} \WHz$, $\ga 10^{21.3} \WHz$,$\ga 10^{22.3} \WHz$, $\ga 10^{23.3} \WHz$, and $\ga 10^{24.3}\WHz$, respectively, assuming a spectral index of $-0.7$. \edit{A value of -0.7-0.8 has commonly been measured~\citep{Smolcic2017a} and typically assumed across literature (e.g. in \citet{Kondapally2022,Whittam2018} and others) }.

Overall, \simba\ predicts rising radio galaxy fractions as a function of stellar mass, qualitatively in accord with observations.  The fractions are lower for brighter $\power$ cuts, and all fractions reach unity at the largest masses. Quantitatively, \simba\ under-predicts the number of radio galaxies with $\power \geq 10^{22} \WHz$ at $10^{11}\la \mstar \msun \la 10^{11.5}$ in comparison to observations. This is not unexpected due to the fact that \simba\ does not produce high luminosity radio galaxies due to the limited $(100\hmpc)^{3}$ volume of the simulation, and are thus missing a large fraction of the radio galaxy population as can be seen by the fraction of sources with $\pmhz \geq 10^{23} \WHz$ in \citet{Sabater2019}. \simba\ produces higher stellar mass radio galaxies than in observations, and predicts that the most massive galaxies with $\mstar \ga 10^{12} \msun$ are radio galaxies. Although we see this offset at intermediate masses, we consider the radio galaxy fractions as a function of $\mstar$ in \simba\ to be a reasonable match to the radio galaxy population.  This is an important check for our environmental study, since higher mass galaxies broadly tend to live in denser environments.

\section{Results}
\subsection{Central and Satellite Radio Galaxies}
\label{sec:cvs}

\begin{figure*}
    \begin{center}
    \includegraphics[width=0.9\textwidth]{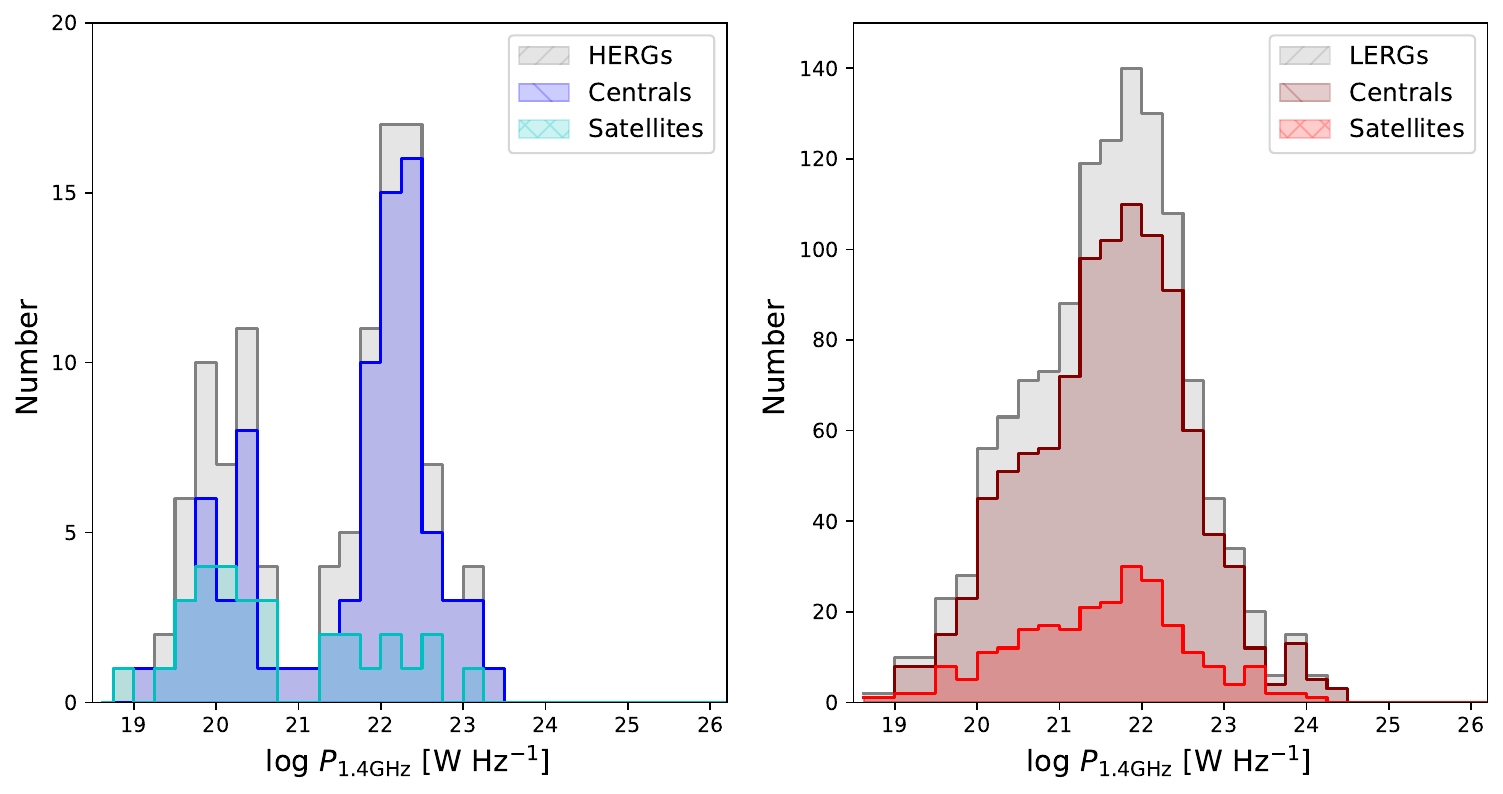}
    \caption{Distributions of 1.4\,GHz radio luminosity for HERGs (grey, left) and LERGs (grey, right), each subdivided into centrals (blue, maroon) and satellites (cyan, red). HERG centrals typically have brighter radio luminosities than HERG satellites, while LERGs show no significant dependence of being a central or satellite on the radio luminosity distribution.}
    \label{fig:Pdist}        
    \end{center}
\end{figure*}

\subsubsection{Satellite fractions}

In \simba, we find that 278 of 1365 radio galaxies are satellites.  This is our first main finding -- while majority of radio galaxies are the central galaxies in their respective halos as canonically believed, 20\% of radio galaxies in \simba\ are in satellite galaxies.  This fraction is consistent with recent observations suggesting a non-trivial minority of bright AGN are hosted by satellites~\citep{Alam2021,Aird2021}. It is worth noting that a large fraction of these satellites have stellar masses comparable to that of their corresponding central, with only 162/278 satellites having masses less that half times the mass of their central and only 77/278 with less than 0.25 times the mass of the central. Galaxy--galaxy mergers are thought to be a significant driver of AGN activity and thus present an interesting question as to how many of these radio galaxy satellites are undergoing mergers, and more generally what role their position in the host halo plays in their evolution. This will be considered in future work, as merger histories are outside the scope of this paper.

Splitting into HERGs and LERGs, in \simba\ 30/113 HERGs are satellites, while 248/1252 LERGs are satellites. This accounts for $26.5\%$ and $19.8\%$ of the respective populations. Hence there is not a large difference in the fractions hosted by satellites for HERGs vs. LERGs.  These fractions can be compared against the total galaxy population, of which 33$\%$ of all resolved galaxies are satellites, while 31$\%$ of galaxies with $\mstar \geq 10^{9.5} \msun$ are satellites.  This shows that radio galaxies are less likely to be hosted by satellites, but not dramatically so.

Hence \simba\ is consistent with the growing evidence that not all radio galaxies are massive central galaxies in groups and clusters.  Next we investigate the differences in the global properties in the populations of central and satellite radio galaxies.

\subsubsection{Radio Luminosities}
Figure~\ref{fig:Pdist} shows the number distribution of the 1.4\,GHz radio luminosity, $\power$, of radio galaxies in \simba. The left panel shows the number distribution of luminosities for HERGs (grey filled) separated into centrals (blue filled) and satellites (cyan filled). The right panel shows the distribution for LERGs (grey filled) separated into centrals (maroon filled) and satellites (red filled).  Table~\ref{tab:ks} shows the $p$-values for a two sample Kolmogorov–Smirnov~(KS) test applied to the fractional histograms of centrals and satellites in terms of $\power$ (note that the histograms shown in Figure~\ref{fig:Pdist} are not fractional, but raw numbers).

In the left panel, we find that more HERG centrals are found at higher radio luminosities \edit{than} HERG satellites. For HERGs, the K-S test finds a $p$-value of $<10^{-3}$, thus rejecting the hypothesis that HERG centrals and satellites are pulled from the same distribution. Since HERGs are defined by the dominant accretion of cold gas, and satellites tend to exist in denser environments than centrals of the same stellar mass, they likely have less cold gas available to be accreted. This results in lower luminosities for HERG satellites versus HERG centrals.

For LERGs, there appears to be no difference in the distribution of radio luminosities of central vs. satellites, which is borne out by the K-S $p\approx 0.2$. Because LERGs are defined as those that are dominant in the accretion of hot gas, whether a LERG is a central or a satellite in a dense environment, the hot gas in this environment is going to dominate the fueling of the central SMBH. 

\begin{figure*}
\begin{center}
    \includegraphics[width=0.9\textwidth]{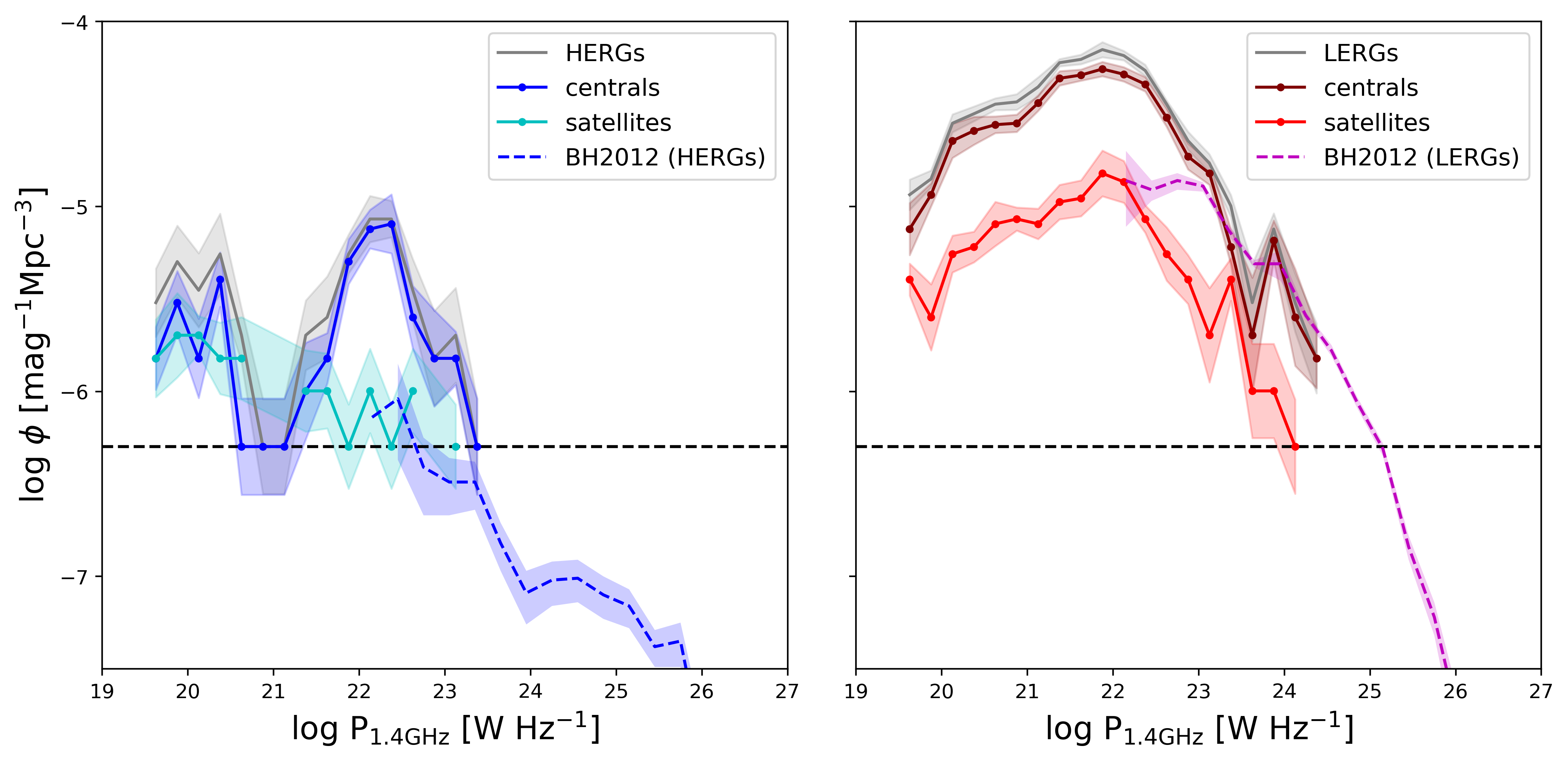}
    \caption{1.4\,GHz radio luminosity functions for HERGs (solid grey, left) and LERGs (solid grey, right) subdivided into centrals (blue, maroon solid with circle markers) and satellites (cyan, red solid with circle markers), in comparison with observations of HERGs and LERGs (blue and magenta dashed) by \citet{BestandHeckman2012}. The horizontal black dashed line shows the number density of one galaxy within the \simba\ volume. For both HERGs and LERGs, centrals dominate the RLF at all luminosities}. 
    \label{fig:RLFs}
\end{center}
\end{figure*}

Figure~\ref{fig:RLFs} provides an alternate view of the relative contributions of central and satellites to the overall radio population and their prevalence per unit volume. We show the 1.4\,GHz radio luminosity functions~(RLFs) with associated 1$\sigma$ bootstrap errors over 50 realisations for, in the left panel, HERGs (grey line) separated into centrals (blue line) and satellites (cyan line) and, in the right panel, LERGs (grey line) separated into centrals (maroon line) and satellites (red line). The RLFs are in comparison to observations of the HERG and LERG RLF by~\citet{BestandHeckman2012} in blue dashed and magenta dashed lines respectively. The black dashed line show the minimum volume density probed in \simba\ due to the limited $(100\hmpc)^{3}$ volume i.e. a single galaxy within the simulation volume).

For HERGs, we see that satellites contributes mostly to the total HERG RLF at low luminosities, although the number of HERGs in the highest luminosity bins are too low to make a statistically significant conclusion. We note a bimodal distribution in the HERG RLF which is consistent with a separation in $sSFR$ (see Figure~\ref{fig:global_props}). Making a direct comparison with observations at these luminosities however remains a challenge due to the volume limit of \simba\ along with the luminosity limit of the observations. These challenges will become less substantial with preliminary results from the MeerKAT International GHz Tiered Extragalactic Exploration~(MIGHTEE,~\citealt{Jarvis2017}) survey which currently covers $\sim$1 deg$^{2}$ of the radio sky down to luminosities of $P_{\rm 1.4\,GHz}\approx 10^{19} \WHz$.
Meanwhile, the contribution of LERG satellites to the LERG RLF is comparable at all luminosities.  These results broadly echo what we saw from the histograms in Figure~\ref{fig:Pdist}.

In summary, we show that  there is no significant difference in the distribution of radio luminosities of central and satellite galaxies. However for HERGs we find that satellites contribute more to the low luminosity end of the distribution of radio luminosities, which we believe is tied to the reduced amounts of cold gas in denser environments, resulting in less efficient accretion onto the black hole.  The LERGs meanwhile show no statistical differences in centrals vs. satellites in terms of their radio power.

\subsubsection{Host Properties}

We here examine the host properties of central and satellite radio galaxies to identify any differences which may relate back to the environmental dependence of black hole growth and feedback. 

\begin{figure*}
\begin{center}
    \includegraphics[width=0.7\textwidth]{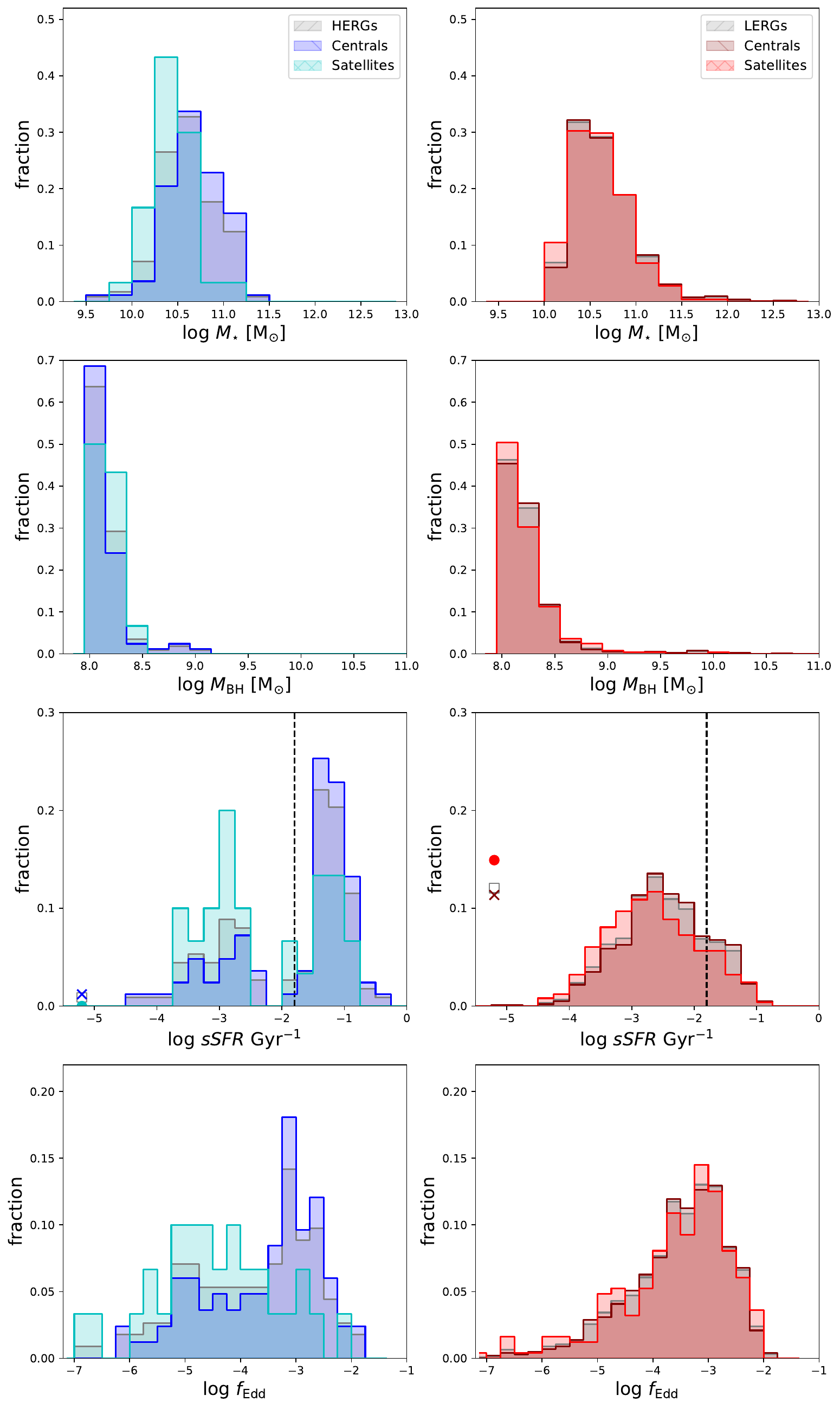}
    \caption{Fractional distributions of global properties of HERGs (left) and LERGs (right) subdivided into centrals (blue, maroon) and satellites (cyan, red). \textit{Top to bottom:} stellar mass $\mstar$, black hole mass $\mbh$, specific star formation rate $sSFR$, Eddington fraction $\fedd$. For $sSFR$, grey squares show the fraction of HERGs or LERGs with $sSFR$s less than $10^{-5}$\,Gyr$^{-1}$. Blue and maroon crosses show the fraction of HERGs and LERGs centrals with $sSFR$s less than $10^{-5}$\,Gyr$^{-1}$. The red filled circle shows the fraction of LERG satellites with $sSFR$s less than $10^{-5}$\,Gyr$^{-1}$, we note that there are no HERG satellites below this threshold. HERG centrals appear to have higher $\mstar$, $\mbh$, and $\fedd$ than HERG satellites, as well as $sSFR$s that span values lower and higher than that of HERG satellites. LERG centrals show higher values of $\mstar$, $\mbh$, and slightly higher values of $sSFR$, while showing no significant differences in the fractional distribution of $\fedd$.}
    \label{fig:global_props}
\end{center}
\end{figure*}

Figure~\ref{fig:global_props} shows the fractional distributions of the global properties of centrals and satellite HERGs (left) and LERGs (right) where the colours are repeated as per Figure~\ref{fig:Pdist}. The properties are, from top to bottom, stellar mass $\mstar$, black hole mass $\mbh$, specific star formation rate $sSFR$, and Eddington fraction $\fedd$. For $sSFR$, the black dashed line shows a separation between star forming and quenched populations at $sSFR=\frac{0.2}{t_{\rm H}}$, where $t_{\rm H}$ is the Hubble time and $sSFR\approx10^{-1.8}$Gyr$^{-1}$ at $z=0$~\citep[e.g.][]{RodriguezMontero2019}. Additionally for $sSFR$, we show the total fraction of galaxies that have $0\leq sSFR/$Gyr\,$\leq 10^{-5}$ as crosses for centrals, filled circles for satellites, and empty grey squares for the total population. \edit{The purpose of which is to visualise these low sSFR sources instead of having them trail off the plot axis.}

For both HERGs and LERGs, we see that the stellar and black hole masses of satellites are slightly smaller than that of centrals. This is qualitatively consistent with the idea that centrals are the highest stellar mass galaxies in their dark matter halo. The $\mbh-\mstar$ relation~\citep[e.g.][]{Thomas2019} additionally implies that the highest $\mstar$ galaxies will also have the largest $\mbh$.

For HERGs that are dominated by the accretion of cold gas onto the central SMBH, the $sSFR$ for satellites span a similar range as that of the centrals. That said, the fraction of satellites at low $sSFR$s appear to be higher than that of centrals, while the contrary holds at higher $sSFR$s. This is due to the fact that richer and hotter environments suppress the accretion of cold gas into satellites, while simultaneously inhibiting cold gas already in the host galaxy from condensing and forming stars, resulting in lower star formation rates. Similarly, the $\fedd$ for satellites are shifted to lower values, peaking at $\fedd \sim10^{-4.5}$, relative to centrals that peak at $\fedd\sim10^{-3}$. This is consistent with the previously stated argument that satellites are limited by the amounts of cold gas that can be efficiently accreted onto the SMBH, resulting in low Eddington fractions.

For LERGs that are dominated by the accretion of hot gas, there is only a minuscule shift in the $sSFR$ distribution of satellites toward lower $sSFR$s than that of centrals. Although LERG satellites have higher star forming gas fractions at the highest environment densities (described later in \S\ref{sec:coldgas}), something is stopping this gas from collapsing, thereby slowing the star formation rate. This may be due to the turbulent hot gas in the circum-galactic medium which, for LERG centrals at the same densities, reduces the cold gas fraction entirely. Additionally, LERGs that are isolated, and thus considered centrals, have high gas fractions and contribute to the high end of the $sSFR$ distribution. There moreover appears to be no significant difference in the distribution of Eddington fractions of central and satellite LERGs, as they are both fuelled by hot gas and are thus not affected by the presence or lack of cold gas.

To quantify our results we again compute the two-sample Kolmogorov-Smirnov~(KS) tests and record the associated $p$-values in Table~\ref{tab:ks}. From top to bottom we consider radio luminosity $\power$ (which we discussed in the previous section), stellar mass $\mstar$, black hole mass $\mbh$, specific star formation rate $sSFR$, and Eddington fraction $\fedd$. The first column shows the associated p-values for HERGs and the second column shows the p-values for LERGs.

\begin{table}
\begin{center}
\begin{tabular}{l c c}
     & HERGs & LERGs\\
\hline
log$\power$ & 0.00076 & 0.20049 \\
log$\mstar$ & 0.00365 & 0.36526 \\
log$\mbh$  & 0.18697  & 0.32450 \\
log $sSFR$     & 0.04764 & 0.00004\\
log $\fedd$ & 0.00076 & 0.79737\\
\hline
\end{tabular}
\captionof{table}{The associated $p$-values recorded from a two-sample Kolmogorov-Smirnov test between populations of central and satellite radio galaxies.}
\label{tab:ks}
\end{center}
\end{table}

For the HERG population, we find the biggest separation in the distributions of $\power$ and $\fedd$ with $p$-values $\ll$ 1$\%$, indicating that HERG satellites accrete less efficiently than HERG centrals. The $\mstar$ distributions for HERGs also show a $p$-value less than 1$\%$, but this is likely due to the definition of centrals in that they have higher $\mstar$. For the $sSFR$ and $\mbh$ distributions, the $p$-values are larger than 1$\%$ but not dramatically so, thus we cannot conclusively state whether the central and satellite populations are drawn from the same distributions.

For the LERGs we find high p-values across the series of host properties, with the notable exception for the distribution of $sSFR$s. As above mentioned and studied further in the next section, the low $sSFR$ satellites are likely to be hosted in more dense environments than that of some potentially more isolated centrals with ongoing star formation.  Thus this makes sense in terms of the way star formation proceeds within halos.

\subsubsection{Cold Gas Properties}
\label{sec:coldgas}
Tracing the cold gas in the hosts of radio galaxies can be used as an indicator of the nature of the gas accretion onto the central SMBH. It is not far-fetched to consider that a HERG-type radio galaxy will have more cold gas in the host galaxy than that of a LERG. We may additionally expect that in higher density environments, the amount and condensation of cold gas is reduced, consequently suppressing the formation of stars~\citep{Hale2018}. 

\begin{figure}
\begin{center}
    \includegraphics[width=0.9\columnwidth]{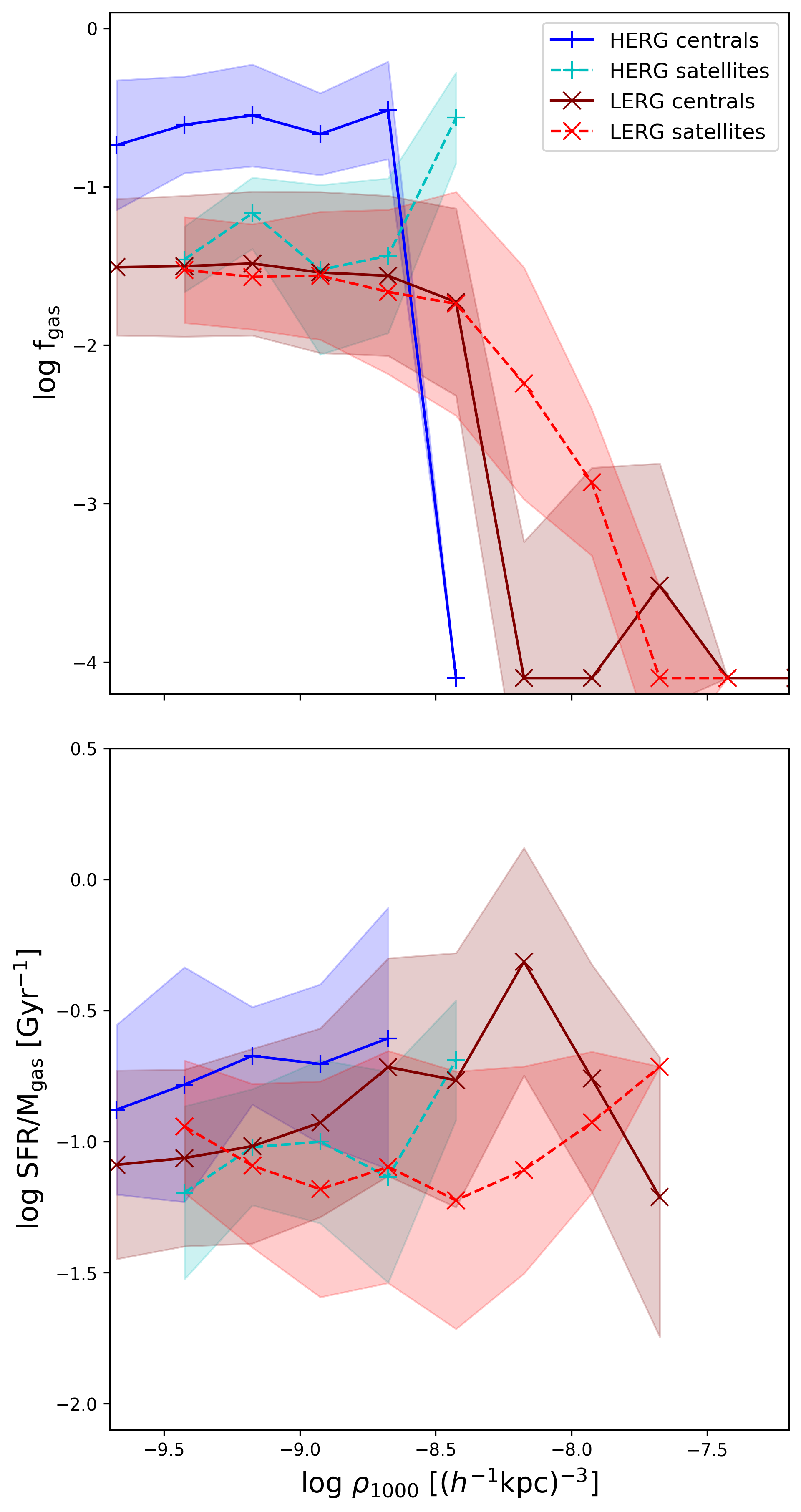}
    \caption{\textit{Top:} Median cold gas fractions for radio galaxies as a function of the density of their environments within a sphere of radius 1000$h^{-1}$kpc. \textit{Bottom:} Median efficiencies of star formation from cold gas as a function of environment density.   
    Blue (+) and maroon (x) sold lines show the fractions for HERG and LERG centrals respectively while cyan (+) and red (x) dashed lines show that of HERG and LERG satellites respectively. Shaded bands show the 1$\sigma$ standard deviation uncertainty.
    At low densities, HERG satellites have lower cold gas fractions than that of HERG centrals but have higher gas fractions at higher densities. LERG centrals and satellites have similar gas fractions at low densities, but LERG centrals drop to little to no cold gas at high densities while LERG satellites still have some cold gas. The efficiency of forming stars from cold gas is higher in both HERG and LERG centrals than in their satellite counterparts. }
    \label{fig:fgas}
\end{center}
\end{figure}

Figure~\ref{fig:fgas} shows the amount of the cold gas in radio galaxies as a function of the galaxy density within a sphere of radius \edit{1000 $h^{-1}$kpc, $\rho_{1000}$}. The top panel shows the median cold gas fractions while the bottom panel shows the ratio of the star formation rate to the gas mass $M_{\rm gas}$, i.e. the star formation efficiency. The blue solid lines with + markers shows the gas fractions for HERG centrals, the dashed cyan lines with + markers shows the gas fractions for HERG satellites. Similarly, the solid maroon lines with x markers show the gas fractions for LERG centrals while the red dashed lines with x markers show that of the LERG satellites. The associated shaded band shows the 1$\sigma$ standard deviation on the median. Galaxies with little to no cold gas, or $\fgas<10^{-4}$, are displayed at $\fgas=10^{-4.1}$..

At low densities or $\rho_{\rm 1000} < 10^{-8.5}(h^{-1}$kpc$)^{-3}$ ($\lesssim13$ galaxies within a $R=1 h^{-1}$Mpc sphere), HERG centrals have higher gas fractions than that of HERG satellites. At the same scales, LERG centrals and satellites have similar gas fractions. At higher densities, the gas fractions for LERG centrals drop to roughly zero while LERG satellites still have some cold gas. When we look at the star formation efficiency, we see that HERG centrals are more efficient at converting their cold gas to stars compared to HERG satellites. Similarly, while LERG centrals have very little cold gas, they are more efficient at converting that cold gas to stars compared to LERG satellites that have a higher fraction of cold gas. This indicates that while the satellite population has significant cold gas reservoirs, they are being inhibited from forming this cold gas into stars, supporting the lower $sSFR$s seen in LERG satellites in Figure~\ref{fig:global_props}.

In summary, there are small differences in the host properties of HERGs depending on whether they are the central or a satellite galaxy in their host dark matter halo, and no statistically significant difference for LERGs except in their $sSFR$.  HERG satellites display slightly lower $sSFR$s, $\fedd$, and $\power$, owing to the reduced amounts of cold gas and turbulent nature of richer environments. Low mass centrals in under dense environments may be the reason for the presence of significant ongoing star formation in the LERG population.  While there are small differences in the host properties of HERGs, there is no strong evidence that being a central or BCG is a required feature in the evolution of radio galaxies. 

\subsection{Clustering of Radio Galaxies}
\label{sec:env}

From here we quantify the environments radio galaxies inhabit, and investigate the role the environment plays in determining global properties of radio galaxies.

\subsubsection{Halo Occupancy Distributions}
\label{sec:hod}
\begin{figure}
    \centering
    \includegraphics[width=\columnwidth]{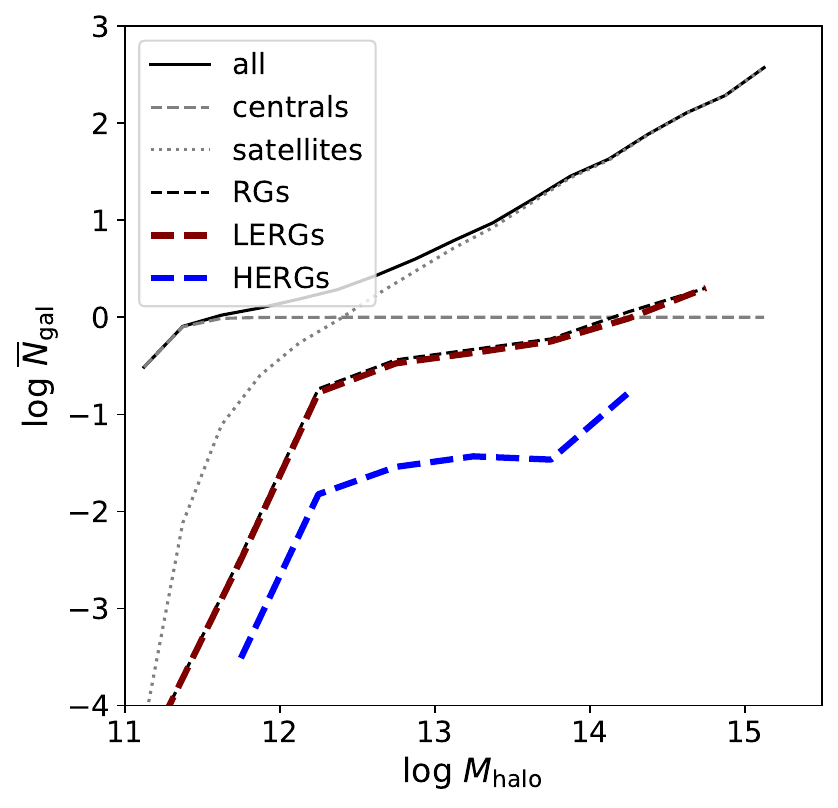}
    \caption{Halo Occupancy Distribution (HOD) for dark matter halos in \simba. Solid black line shows the HOD for all resolved galaxies, subdivided into central galaxies (grey dashed) and satellite galaxies (grey dotted). The HOD for radio galaxies is shown by the dashed black line, subdivided into HERGs (blue dashed) and LERGs (maroon dashed). The most massive halos can host more than one LERG, while HERGs tend to reside in lower mass halos that do not host other HERGs.}
    \label{fig:HOD}
\end{figure}

Figure~\ref{fig:HOD} shows the Halo Occupancy Distribution~(HOD) for galaxies in \simba. The HOD describes the average number of galaxies in a given dark matter halo of mass $\mhalo$ and acts as an estimate of the underlying dark matter distribution for observed isolated, grouped, or clustered galaxies. The black line shows the HOD for all resolved galaxies in \simba, while the dashed and dotted grey lines show the HOD for \simba\ galaxies separated into centrals and satellites respectively. 

As expected, the average number of centrals converges to 1 at all halo masses. This is due to the fact that halos can only host one central at any given time by definition, that is, that the central has the highest stellar mass in a given dark matter halo or, as in observations, is the galaxy closest to the gravitational potential minimum of the group or cluster. At low halo masses, the average number of centrals decreases due to the presence of dark matter halos containing no or only unresolved galaxies owing to our galaxy mass resolution limit. The average number of satellites also decrease toward low halo masses due to the decrease in the total number of galaxies for those halos, while the most massive halos with $\mhalo \ga 10^{14.5}\msun$ can host hundreds of galaxies by z=0.
Satellites dominate the HOD at $\mhalo \ga 10^{12.4} \msun$ in comparison with $\mhalo \ga 10^{13} \msun$ in observations of DEEP2 and SDSS galaxies~\citep{Zheng2007} however, these observations are flux-limited so may miss some of the low mass, low luminosity satellites in lower mass halos. \edit{Similarly, low mass ($\mstar<5.8\times10^{8}\msun$) satellites are not resolved in \simba\ and we therefore emphasise that this comparison is more suited to flux limited observations of more massive galaxies. }

The black dashed line shows the HOD for all radio galaxies, while the maroon and blue dashed lines shows the HOD for LERGs and HERGs respectively. The average number of radio galaxies increase with halo mass, which is expected since more massive halos typically host more massive galaxies according to the $\mstar-\mhalo$ relation, and since radio galaxies have $\mstar >10^{9.5} \msun$ in \simba. Note that this $\mstar-\mhalo$ relation is well established for centrals which dominate the radio galaxy population, but is weaker for satellites. At $\mhalo>10^{14}\msun$, of which there are 43 dark matter halos in \simba, there is on average 1 or more LERGs per halo.  HERGs in \simba\ are not found in halos with $\mhalo>10^{14.3}\msun$, while LERGs can be found in more massive halos. \edit{\simba's limited volume and hot mode accretion dominated radio galaxies in high mass halos may therefore be responsible for the lack of HERGs seen at high halo masses.} Both populations drop of sharply at $\mhalo\la 10^{12.2}\msun$, owing to our threshold of $\mstar \geq 10^{9.5} \msun$ for radio galaxies.
Our results are in agreement with the hypothesis that LERGs live in more dense environments than HERGs.  These results agree with~\citet{Croston2019} who found that only a very small fraction of AGN reside in clusters with $\mhalo > 10^{14} \msun$, and in addition show that at these same high masses, halos can host more than one LERG at $z=0$. However, because there are significantly fewer radio galaxies in \simba\ than non-radio galaxies, counting galaxies alone via an HOD cannot allow us to compare radio galaxy environments to that of the overall galaxy population.  We next consider other environment probes to explore this in more detail.

\subsubsection{Two Point Correlation Functions}
The two point correlation function~(TPCF) compares the clustering properties of galaxies by measuring the excess probability of finding two galaxies at a given separation. In other words, it defines how strongly and at which scales galaxies cluster together, compared to a universe with a uniform distribution of galaxies. We know that the galaxies in our universe are not uniformly distributed as the $\Lambda$CDM model describes the formation of the Cosmic Web of filaments, sheets, and nodes. We expect the clustering strength of galaxies to drop off at scales larger than the size of their halo which is typically on $\sim$Mpc scales. Because massive halos host many massive galaxies, we expect the most massive galaxies to cluster more strongly than smaller galaxies. A corollary of this is that passive galaxies cluster more strongly than star forming galaxies. Considering that radio galaxies are hosted typically by massive elliptical galaxies, we expect that radio galaxies cluster more strongly than SFGs, as seen in~\citet{Hale2018}.  Here we quantify these trends in \simba.

We compute the 3D TPCF using:
\begin{equation}
    \xi_{g} = \frac{DD-RR}{RR}.
\end{equation}
Where DD is the number of pairs of galaxies in \simba\ in a given radial shell, and RR is the analytically-computed number of expected galaxy pairs in a given radial shell assuming a homogeneous distribution of galaxies, computed as 
\begin{equation}
    RR = \rho \times \frac{4}{3} [(r+\delta r)^3 - r^3],
\end{equation}
where $\rho$ is the number density in a radial shell from $r\to r+\delta r$.

\begin{figure}
    \centering
    \includegraphics[width=\columnwidth]{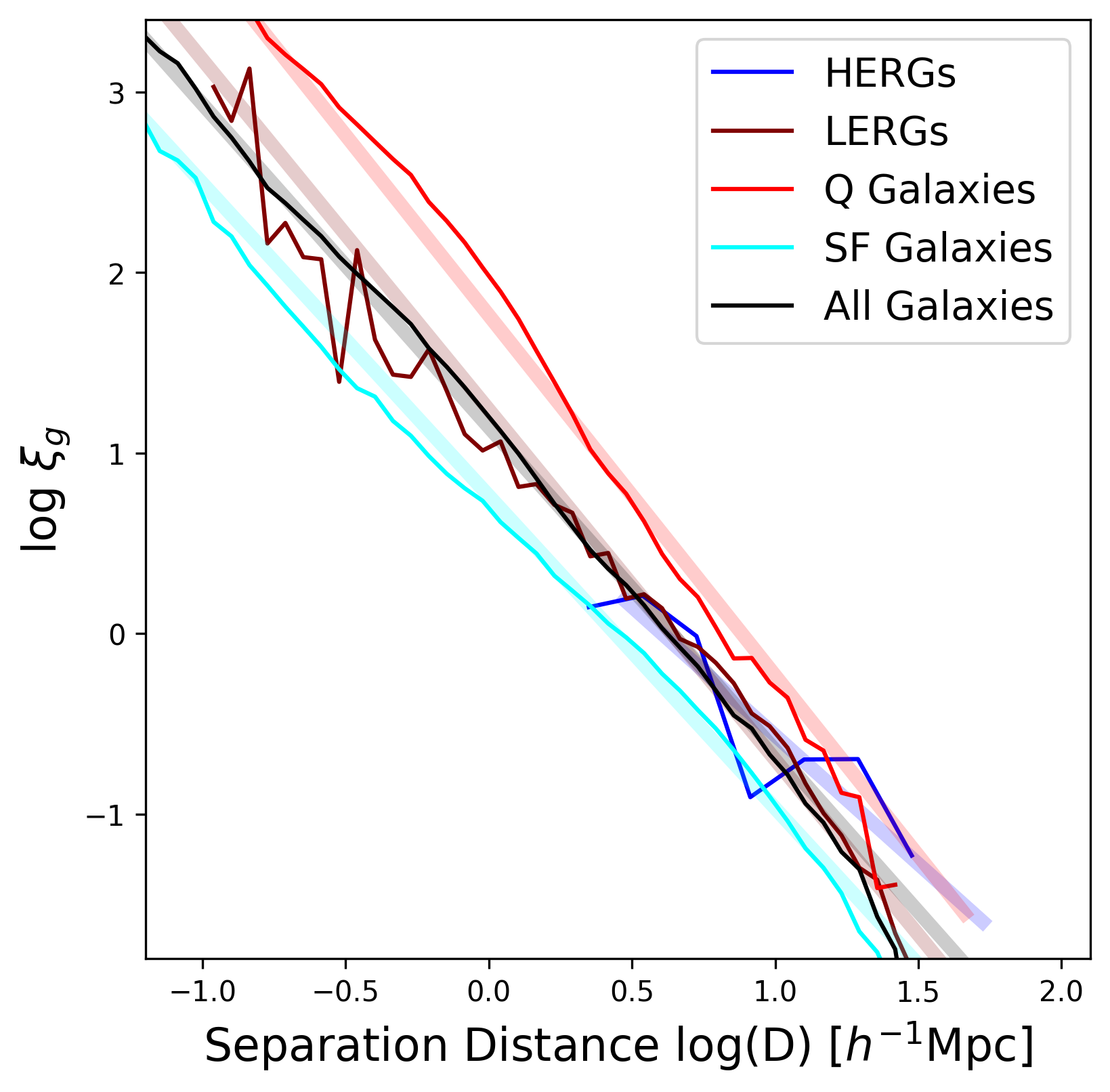}
    \caption{The 3D Two Point Correlation Functions~(TPCFs) for LERGs (maroon line), and HERGs (blue line) in \simba\ at $z=0$. We show also for reference the TPCF for the total galaxy population in \simba~(All, black line) split into quiescent (Q, red line) and star forming (SF, cyan line) galaxies. The thick lines with low opacity are the best fit power law to the corresponding TPCFs. LERGs cluster similarly to the total resolved galaxy sample while HERGs do not cluster at small scales indicating they are unlikely to be found within the same dark matter halo.}
    \label{fig:tpcf}
\end{figure}

Figure~\ref{fig:tpcf} shows the 3D spatial TPCF for various galaxy populations in \simba. The maroon line shows the TPCF for LERGs, while the blue line shows that for HERGs. We show additionally the TPCF for total galaxy population in \simba\ represented by the black line and which is split into quiescent galaxies in red and star forming galaxies in cyan. The thick lines with low opacity represent the best fit power law to the colour-corresponding TPCFs. The power law is in the form $\xi = (\frac{r_{0}}{r})^{\gamma}$, where $\gamma$ is the slope of the TPCF and $r_0$ is the clustering length. We show the associated $r_0$ and $\gamma$ values in Table~\ref{tab:tpcf_params}.

The total (resolved) galaxy population has a clustering length of $r_0 = 3.86h^{-1}$Mpc. Quiescent galaxies cluster the most strongly showing the highest excess in number compared to that of a uniformed distribution of quiescent galaxies with galaxy pairs separated by distances up to 10$\hmpc$ and with a cluster length of $r_0 = 6.79h^{-1}$Mpc and $\gamma=-2.12$, which is similar to observations of red galaxy clustering~\citep{Coil2017}. Additionally, the steep slope in the quiescent galaxy TPCF indicates that the excess in galaxy pairs on small scales is much higher than on larger scales, indicating that quiescent galaxies typically cluster much more strongly on small scales in galaxy groups and clusters. In contrast, the excess in galaxies pairs across all scales (i.e. the clustering strength) of star forming galaxies is much weaker than other galaxies, with a clustering length of $r_0 = 2.37h^{-1}$Mpc and $\gamma=-1.63$. 

\begin{table}
\begin{center}
\begin{tabular}{l c c}
     & $r_0$ ($h^{-1}$Mpc) & $\gamma$\\
\hline
HERGs & 2.99 & -0.95 \\
LERGs &  4.08 & -1.59 \\
Q  &  6.79 &  -2.12\\
SF     & 2.37 & -1.63\\
All & 3.86 & -1.80 \\
\hline
\end{tabular}
\captionof{table}{The associated $r_0$ and $\gamma$ values best fit to the TPCF of HERGs, LERGs, quiescent galaxies (Q), star forming galaxies (SF), and the total resolved galaxy population (All) in \simba .}
\label{tab:tpcf_params}
\end{center}
\end{table}

Shifting our focus to radio galaxies, The population of LERGs show a clustering length of $r_0 = 4.08h^{-1}$Mpc and $\gamma=-1.59$ and similar to that of the total galaxy population, implying LERGs follow the same clustering probability than the total galaxy population.  While HERGs may have a clustering length of $r_0= 2.99h^{-1}$Mpc and $\gamma=-0.95$, the population does not cluster on small scales but clusters on scales of the size of the associated halos and thus similarly to the two-halo term of the total TPCF~\citep{Zheng2007}. This indicates that multiple HERGs are not typically found in the same dark matter halo, echoing the result found in \S\ref{sec:hod} where the average number of HERGs in a dark matter halo at all masses is less than 1. \edit{This is reiterated and confirmed in that, upon inspection, no single dark matter halo in \simba's $100\hmpc$ cubed volume hosts more than 1 HERG. A consequence of selecting radio galaxies as defined in \S\ref{sec:radio_em} results in a population of hosts that are less likely to be fueled by cold mode accretion. This is observed by the small fraction of HERGs i.e. 113/1365 radio galaxies. Additionally, cold mode accretors are likely to be found in low density environments as we will see later in \S\ref{sec:environment} and \ref{sec:env_dense}. It is therefore not surprising that we do not find any 2 HERGs hosted by the same dark matter halo. However, whether this is a consequence purely of the HERG/LERG definition used throughout this work or that of \simba's limited volume poses an interesting prediction for cluster and radio surveys.}.

In summary, LERGs cluster similarly to the total galaxy population, stronger than star forming galaxies and less strongly than quiescent galaxies.  HERGs however cluster only at scales larger than the typical host dark matter halo which implies a low probability of finding more than one HERGs within the same dark matter halo. These results show that there is no exaggerated bias in the clustering of LERGs, while we do not expect high clustering probability for HERGs at short scales.  With upcoming wide-area multi-wavelength surveys, these predictions provide an interesting test of how \simba\ populates radio galaxies within large-scale structure.

\subsubsection{Environmental Richness}
\label{sec:environment}

Another way to characterise the environment of radio galaxies is to examine their environmental density, to understand their typical location in the Cosmic Web.
To explore the host environments of radio galaxies and their physical properties, we consider the environmental richness and local density as a function of host galaxy and black hole properties.  We define the richness parameter as the number of galaxies residing in the host dark matter halo, and we will examine richness as a function of their $\mstar$, $\mbh/\mstar$, $\power$, and $\fedd$ for radio vs. non-radio galaxies with $\mstar\geq10^{9.5}\msun$.

\begin{figure*}
\begin{center}
    \includegraphics[width=\textwidth]{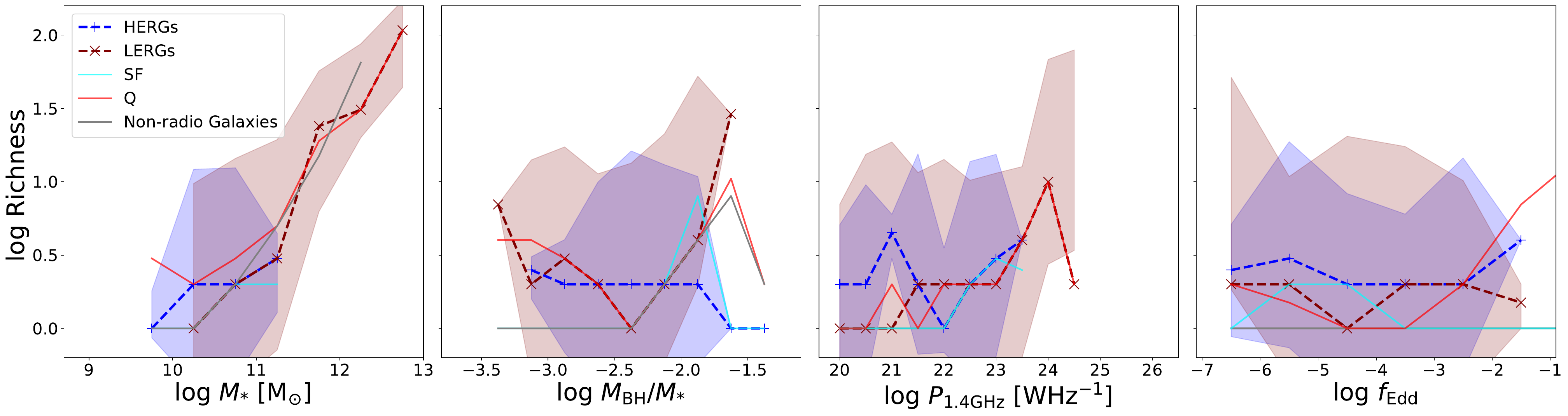}
    \caption{Richness of the environments of radio galaxies as a function of stellar mass $\mstar$, specific black hole mass $\mbh/\mstar$, 1.4 GHz radio luminosity $\power$, and Eddington fraction $\fedd$. Maroon dashed lines show the richness of the environments for LERGs, the blue dashed lines shows that for HERGs, and the grey solid lines is that for all non-radio galaxies in \simba. Additionally the total galaxy population is split into star forming and quiescent galaxies, shown by the cyan and red lines respectively. The red and blue bands show the 1$\sigma$ standard deviation on the median errors for LERGs and HERGs respectively. The richness of the environments of radio galaxies correlates more strongly with stellar mass and over-massive black hole hosts. The richness of different radio galaxies' environments are similar across $\power$ and $\fedd$.}
    \label{fig:richness}
\end{center}
\end{figure*}

\begin{figure*}
\begin{center}
    \includegraphics[width=\textwidth]{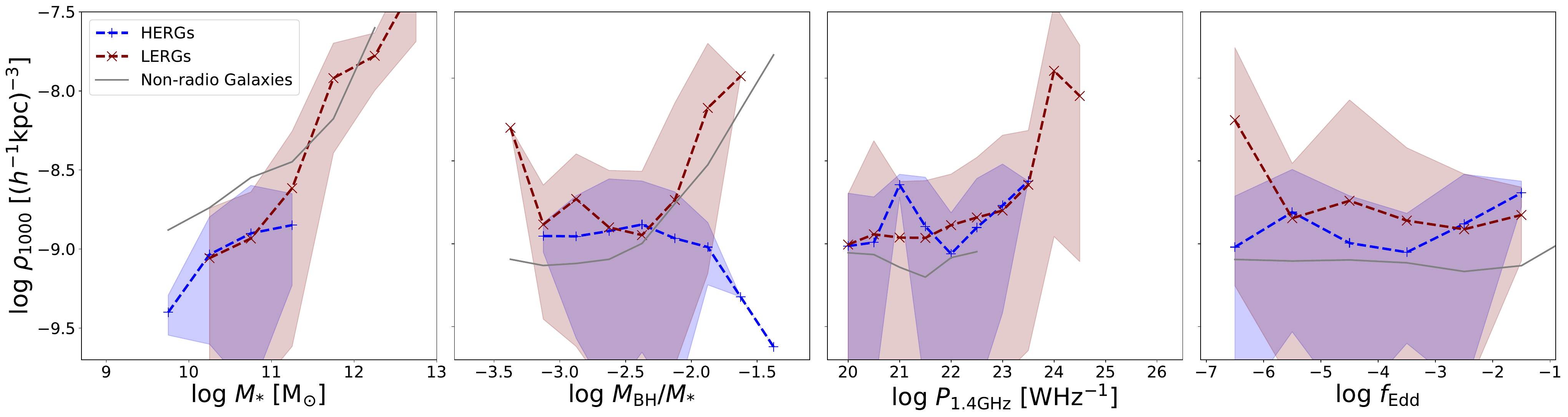}
    \caption{Density of the environments of radio galaxies in a sphere of radius 1000 $\hkpc$ as a function of stellar mass $\mstar$, specific black hole mass $\mbh/\mstar$, 1.4 GHz radio luminosity $\power$, and Eddington fraction $\fedd$. Maroon dashed lines show the richness of the environments for LERGs, the blue dashed lines shows that for HERGs, and the grey solid lines is that for all non-radio galaxies in \simba. The red and blue bands show the 1$\sigma$ standard deviation on the median errors for LERGs and HERGs respectively. The density of the environment of radio galaxies correlates more strongly with stellar mass, and thus high mass LERGs live in more rich environments than HERGs. LERGs with over-massive black holes reside in high density environments while HERGs with over-massive black holes reside in low density environments. There is a slight correlation between density and $\power$ however the density of different radio galaxies' environments are similar across $\power$ and $\fedd$.}
    \label{fig:density}
\end{center}
\end{figure*}

Figure~\ref{fig:richness} shows the running median relations for the richness of the environments of galaxies in \simba. The richness of LERGs and HERGs are shown by maroon and blue dashed lines respectively. Additionally we show the sample of galaxies in \simba\ that are not radio galaxies by the grey line. The full galaxy sample with $\mstar\geq10^{9.5}\msun$ is split into populations of quenched and star forming galaxies and are shown by the red and cyan lines respectively. We show the associated 1$\sigma$ standard deviation on the median for only the HERG and LERG populations so as to avoid clutter. The plots following this will include only the radio galaxy and non-radio galaxy populations for a clearer comparison.

In the first panel we show the richness as a function of stellar mass. At stellar masses of$\mstar\la10^{11}\msun$, radio galaxies are less likely to live in rich environments compared with the non-radio and quenched galaxy populations. At these stellar masses, HERGs and LERGs both follow the richness of the star-forming galaxy population.  As expected, the likelihood of living in a rich environment increases with stellar mass. In particular, at $\mstar \ga 10^{11}\msun$ the likelihood of LERGs living in rich environments increases dramatically. This is supported by the idea that clusters host predominantly quenched galaxies and contain high temperature gas. HERGs do not populate such environments, but the few HERGs at high masses tend to live in similar environments to star-forming galaxies galaxies.

Another way to view these trends is that, at a given richness, radio galaxies tend to live in more massive galaxies, partly a consequence of the mass limit imposed for radio galaxies in \simba.  So if one looks in a given galaxy group, according to \simba, one expects to see radio galaxies arising in the higher stellar mass galaxies. 

The specific black hole mass is defined as the ratio between black hole and stellar mass. Based on the $\mbh-\mstar$ relation found in \citet{Thomas2019} we compute the median specific black hole mass for $10^{8}\leq\mbh/\msun\leq10^{10}$ to be $10^{-2.5}\la\mbh/\mstar\la10^{-2.25}$. We thus consider specific black hole masses outside of these boundaries to be under- or over-massive respectively. For richness vs. specific black hole mass (second panel) there is not nearly as much difference between radio galaxy and quenched populations for $\mbh/\mstar\la10^{-2}$; at a given $\mbh/\mstar$, these populations live in similarly rich environments. While most star-forming galaxies live in isolated environments, those with over-massive black holes reside in more rich environments. These black holes in these rich environments are likely to have a significant fraction of their accretion via hot mode growing the black hole much more rapidly due to the high black hole mass dependence of the Bondi rate. This is supported by the same trend shown in that of quiescent populations at high specific black hole masses. 
The richness of a LERG's environment similarly increases, however more rapidly toward sub-populations with over-massive black holes compared to all other populations.  This indicates that the hot circum-galactic medium is key in the fuelling of LERGs via hot mode accretion. The richness of a HERGs environment decreases toward high $\mbh/\mstar$, indicating the contrary for LERG populations in that the hot circum-galactic medium inhibits the growth of cold mode accreting HERGs. \edit{\citet{vanSon2019} tracked the evolution of the 5 most over-massive black holes (or ''black hole monster galaxies''~(BMGs)) in the Hydrangea/C-EAGLE simulations and found that these BMGs reach their positions on the $\mbh-\mstar$ relation due to a combination of early formation times and tidal stripping within dense environments. Black holes in these simulations accrete via Bondi accretion similarly to our LERGs and thus poses a question as to whether these processes are significant for our results and the contrary seen for cold-mode accreting HERGs. This is however beyond the scope of this paper at this point and may be considered in future works.}

The relationship between richness and radio luminosity is less strong. At $\power \la 10^{24} \WHz$ the flat slope and large errorbars imply that there is no significant dependence on luminosity in defining the richness of the environment for all galaxy populations, except for a notable increase in richness at $\power \ga 10^{24} \WHz$ populated only by LERGs. This is in agreement with observations that find that only at the highest luminosities ($\power\ga10^{24}\WHz$) is there a relation between the central AGN and the properties of the group or cluster~\citep{Ching2017}. 

Similarly there is no strong overall dependence of Eddington fraction on richness in all populations. This relation is additionally weakened by the large uncertainties which arise due to the large spread in $\power$ and $\fedd$ as seen in Figures~\ref{fig:Pdist}-\ref{fig:global_props}.
We note that for galaxies to have a non-zero Eddington fraction, the central SMBH is required to have a non-zero accretion rate. This requirement implies that we are not directly comparing to star forming, quiescent, or non-radio galaxies, but to active black holes within those populations that lack kinetic jets. If we consider the richness as a function of $\fedd$, HERGs and LERGs live in richer environments than non-radio galaxies with accreting black holes, as well as star forming galaxies with accreting black holes at all $\fedd$. Alternatively, from this we note that we do not expect to find accreting black holes in non-radio or star forming galaxies in groups or clusters with three or more galaxies, additionally indicating that these galaxies typically live in isolated environments or pairs. 

\subsubsection{Environmental Density}
\label{sec:env_dense}
To quantify the environment in a halo-independent manner and avoid any direct bias owing to the size of the halo, we investigate the surrounding density of a galaxy, $\rho_{\rm 1000}$, by defining a 1000$\hkpc$ tophat sphere around the galaxy and determining the local number density within that sphere.

Figure~\ref{fig:density} shows the running median relations for $\rho_{\rm 1000}$ as a function of $\mstar$, $\mbh/\mstar$, $\power$, and $\fedd$. We show only the populations of HERGs (blue dashed lines) and LERGs (maroon dashed lines) and compare this with the non-radio galaxy population (grey solid line). For clarity, we show the 1$\sigma$ standard deviation on the median uncertainties for the radio galaxy populations only.

We find overall similar trends to that of the richness of the galaxy populations halos.  HERGs reside in less dense environments compared to that of the overall non-radio galaxy population at the same stellar mass, similar to star-forming galaxies. Low-$\mstar$ LERGs reside in less dense regions as well, however the more massive LERGs end up in denser environments, comparable to non-radio galaxies; in large part this is because, as shown in Figure~\ref{fig:fagn}, most very massive galaxies host radio galaxies (which are uniformly LERGs).

The environment density of LERGs and non-radio galaxies similarly increases strongly with specific black hole mass whilst decreasing for HERGs.  However, at the same value of specific black hole mass for $\mbh/\mstar\la10^{-2.4}$ there is no difference between the environments of the radio and non-radio galaxy populations. This supports the idea that high density environments are key in supporting hot gas accretion onto black holes while inhibiting the efficient accretion of cold gas which dominates HERGs and a fraction of the non-radio galaxy population. In addition, and similarly to what was seen for richness, there is no strong or robust dependence in density on radio luminosity, other than a very slight increase toward high luminosities culminated by a spike in density at luminosities of$\power\ga10^{23.5}\WHz$. At a given luminosity, there is no difference in the environments of HERGs or LERGs. Furthermore, radio galaxies reside in more dense regions than non-radio galaxies that have a radio luminosity due to ongoing star formation processes.
There is essentially no trend with $\fedd$, except at the very lowest $\fedd$ values, the LERGs live in denser-than-typical regions while HERGs live in less dense regions, and radio galaxies live in more dense environments than non-radio galaxies with accreting black holes across $\fedd$.

\begin{figure*}
\begin{center}
    \includegraphics[width=\textwidth]{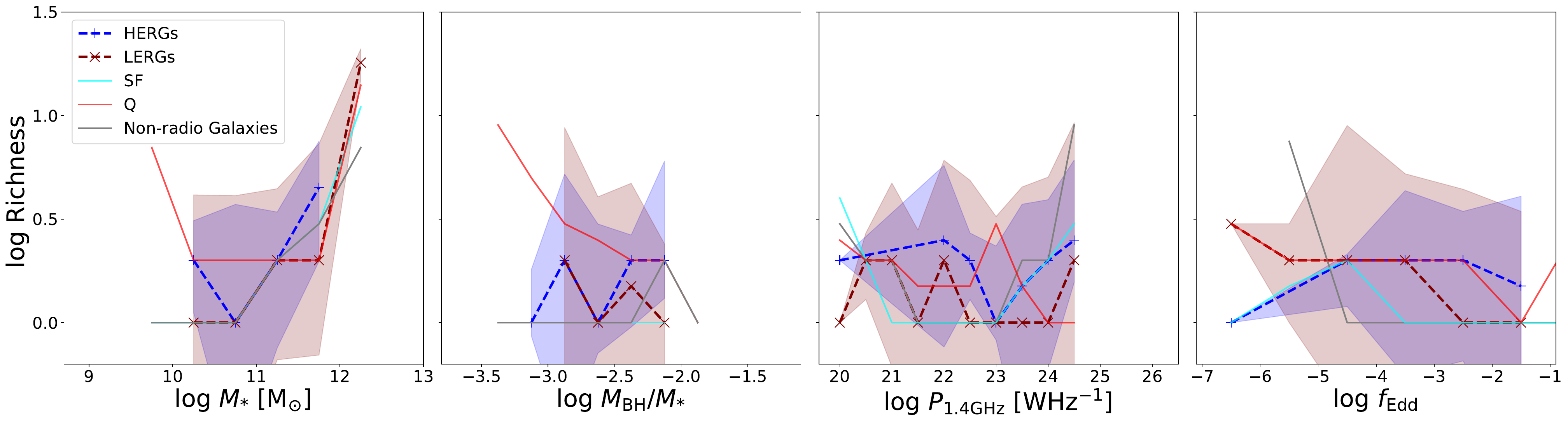}
    \caption{Richness of the environments of radio galaxies at $z=2$ as a function of stellar mass $\mstar$, specific black hole mass $\mbh/\mstar$, 1.4 GHz radio luminosity $\power$, and Eddington fraction $\fedd$. Maroon dashed lines show the richness of the environments for LERGs, the blue dashed lines shows that for HERGs, and the grey solid lines is that for all non-radio galaxies in \simba. The red and blue bands show the 1$\sigma$ on the median errors for LERGs and HERGs respectively. The difference in the richness of the environments of HERGs, LERGs, and non-radio galaxies are similar across $\mstar$, $\mbh/\mstar$, and $\power$. At $\fedd\lesssim10^{-4}$ LERGs reside in more rich environments than HERGs. }
    \label{fig:richness_highz}
\end{center}
\end{figure*}

\begin{figure*}
\begin{center}
    \includegraphics[width=\textwidth]{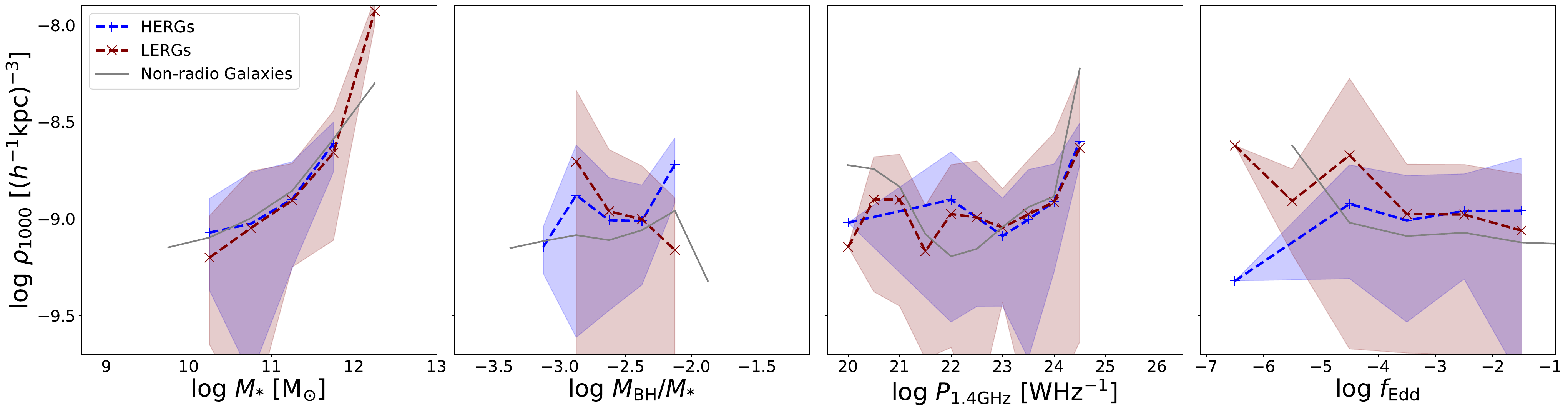}
    \caption{Density of the environments at $z=2$ of radio galaxies in a sphere of radius 1000 $\hkpc$ as a function of stellar mass $\mstar$, specific black hole mass $\mbh/\mstar$, 1.4 GHz radio luminosity $\power$, and Eddington fraction $\fedd$. Maroon dashed lines show the richness of the environments for LERGs, the blue dashed lines shows that for HERGs, and the grey solid lines is that for all non-radio galaxies in \simba. The red and blue bands show the 1$\sigma$ on the median errors for LERGs and HERGs respectively. The difference in the density of the environments of HERGs, LERGs, and non-radio galaxies are similar across $\mstar$, $\mbh/\mstar$, and $\power$. At $\fedd\lesssim10^{-4}$ LERGs reside in more dense environments than HERGs. }
    \label{fig:density_highz}
\end{center}
\end{figure*}

Of the properties considered here, we conclude that the strongest factor that is indicative of the richness of galaxy's environment is the black hole mass, which itself is highly correlated with stellar mass. Specifically, LERGs with higher stellar and black hole masses trace a rich environment while HERGs, as well as lower-mass LERGs, live in less dense environments similar to SFGs.  At the luminosities probed in \simba, we note that there is no strong dependence on the 1.4\,GHz luminosity of radio galaxies on the environment in which they are found.  However, at the same luminosities, radio galaxies may live in slightly more dense environments than non-radio galaxies with ongoing star formation. LERGs with lower $\fedd$ reside in higher density regions compared to HERGs and non-radio galaxies with black holes accreting at the same Eddington rates, as well as to LERGs accreting at higher Eddington rate. 

These results imply that, at the luminosities probed in \simba, the notion of finding radio galaxies in dense groups or cluster-like environments is mostly established by the relationship between halo mass with stellar mass in agreement with conclusions from~\citet{Best2005b}. Similarly, LERGs in the most dense environments have over-massive black holes -- an indication that high density environments support hot mode accretion, while the contrary holds for HERGs with over-massive black holes which reside in low density environments. However, at more moderate masses and lower radio powers, radio galaxies (both HERGs and LERGs) tend to follow the environments of the star-forming galaxy population, which are typically in less dense environments than the total galaxy population. \edit{The uncertainties seen in these trends are significant and so caution must be taken in the interpretation of these results.} These predictions \edit{however} can be \edit{constrained} with upcoming multi-wavelength radio surveys such as MIGHTEE and the MeerKAT Galaxy Cluster Legacy Survey~(MGCLS,~\citealt{Knowles2022}) on the MeerKAT radio telescope which will observe galaxy environments and the role of AGN within cluster evolution.

\subsubsection{High redshift environmental density}

The colour--density relation of galaxies is known to evolve strongly with redshift~\citep{Cooper2007}, and given the connection between radio jets and quenching, it is interesting to examine if the relationship between radio galaxies and density also evolves.  Thus here we investigate the evolution in the environment of radio galaxies at $z=2$, to contrast with our previous results at $z=0$.

Figure~\ref{fig:richness_highz} shows the running median relations for the richness of the environments of radio and non-radio galaxies at $z=2$ and it's dependence of $\mstar$, $\mbh$, $\power$, and $\fedd$.  Populations are represented in the same manner as \S\ref{sec:environment}, that is, LERGs are represented by the maroon dashed line, HERGs are represented by the blue dashed line, and both show associated 1$\sigma$ uncertainties. Non-radio galaxies are shown by the solid grey line.

We note that the densities of halos at $z=2$ are much lower due to having fewer massive galaxies formed and accreted into dark matter halos within the simulation at this time with a total of 136 HERGs and 249 LERGs. However, we notice similar trends to $z=0$ in that richness most strongly depends on stellar mass. A notable difference is that, at the same stellar masses, there is no difference in the halo environment of radio and non-radio galaxies. 

In \simba\ there is no significant evolution in the $\mbh-\mstar$ relation~\citep{Thomas2019} thus the same limits for under- and over-massive black holes exist at $z=2$ as at $z=0$. It is worth noting that, at $z=2$ however, there are no significantly over-massive black hole hosts. At all specific black hole masses, there are no significant differences in the richness of the environments of HERGs and LERGs nor is there any correlation between the specific black hole mass of a radio galaxy and the richness of their environment, while for the quiescent population, galaxies with under-massive black holes reside in more rich environments than quiescent galaxies on the median $\mbh-\mstar$ relation. These are likely satellite galaxies that have been recently accreted into their host dark matter halo.

Another trend similar to that seen at $z=0$ is that there is no strong dependence of richness on radio luminosity. However, radio galaxies of the same luminosity as non-radio galaxies with ongoing star formation are more likely to reside in more rich halos.

There is no strong correlation between richness and $\fedd$. At low $\fedd$, HERGs reside in lower richness environments compared to that of LERGs, star forming, and quiescent galaxies. The scatter in this, however, would make any correlation unreliable. Therefore, at the same $\fedd$ radio galaxies and the few non-radio galaxies with accreting SMBHs reside in similarly rich environments.

We find similar results when investing the environmental density at $z=2$ in Figure~\ref{fig:density_highz}. There is a strong dependence of the local density of galaxies based on stellar mass. Here we see that at low stellar masses, LERGs reside in mildly less dense environments than HERGs and non-radio galaxies of the same stellar mass. Radio galaxies live in mildly higher density environments than non-radio galaxies at intermediate $\mbh/\mstar$. The median relations between $\mbh/\mstar$ and density is anti-correlated for LERGs while mildly positively correlated for HERGs which, although there are significant uncertainties associated with these correlations, is significantly different to the trend we see at $z=0$ for $\mbh/\mstar\ga10^{-2.4}$. This is tied to the evolution of gas content toward higher redshift, in particular that galaxies contain more cold gas~\citep{Dave2020}. Additionally in \simba, the cosmic black hole accretion rate density follows that of the cosmic star formation rate density, both peaking roughly before $z\sim 2$~\citep{Thomas2019}. The high accretion rates and high content of cold gas allows for cold mode black holes to grow more efficiently~\citep{Thomas2021} in conjunction with the large amounts of cold gas in high density regions that have not yet been heated and ionised.  

The same trend seen in richness is seen for radio luminosities as well as Eddington fractions. At intermediate luminosities of $10^{21}\la\power/\WHz\la10^{23}$, radio galaxies reside in more dense environments than non-radio galaxies with radio luminosities due to on-going star formation. However at the same $\fedd$, HERGs, LERGs, and non-radio galaxies with accreting SMBHs all live within similar environments. \edit{As in the previous sub-section, we emphasise the significance of the uncertainties in these trends.}

In summary, at $z=2$ as at $z=0$, density is most strongly correlated with stellar mass.  However radio and non-radio populations at the same $\mstar$, $\power$, or $\fedd$ reside in similar density environments, with a very minor correlation between the type of galaxy population and their environment.  At Cosmic Noon, \simba\ predicts that high density environments are more efficient in supporting the growth of HERGs due to higher availability of cold gas and higher cosmic accretion rate densities. 

\section{Conclusions}
\label{sec:conclude}

We use the state-of-the-art \simba\ suite of cosmological hydrodynamic simulations to study the relationship between radio galaxies and their environments at $z=0$ and how this relationship affects their evolution. Using a population of radio galaxies defined in~\citet{Thomas2021} and separated into high excitation radio galaxies~(HERGs) and low excitation radio galaxies~(LERGs) based on their accretion being dominated by cold or hot mode respectively, we investigate the properties of HERGs and LERGs based on whether they are identified as central or satellite galaxies in their host dark matter halo. The properties we consider are the 1.4\,GHz radio luminosity $\power$, stellar mass $\mstar$, black hole mass $\mbh$, specific star formation rate $sSFR$, Eddington fraction $\fedd$, and cold gas fractions $\fgas$. In addition we study the clustering properties of radio galaxies and consider the environments these objects are likely to be found in, and check how this changes from $z=2$. By classifying our radio galaxy population into centrals and satellites, we find the following results: 
\begin{enumerate}[i]
    \item $\sim20\%$ of radio galaxies are satellite galaxies, accounting for $\sim27\%$ of HERGs and $\sim19\%$ of LERGs. Compared to 33$\%$ of all resolved galaxies in \simba\ and 31$\%$ of all galaxies with $\mstar \geq 10^{9.5} \msun$, radio galaxies avoid satellite galaxies but not dramatically. The radio galaxies identified as satellites show lower stellar and black hole masses compared to central radio galaxies. This is a consequence of the definition that centrals are the most massive galaxy in their dark matter halo.
    \item The $\power$, $sSFR$s, and $\fedd$ of LERGs do not significantly vary whether identified as a central or satellite. Meanwhile, HERG satellites have lower luminosity than HERG centrals and the hosts of HERG satellites show lower $sSFR$s and $\fedd$ than HERG centrals.
    \item At low density environments LERG centrals and satellites have similar gas fractions. In contrast, in high density environments LERG centrals have little to no cold gas, while LERG satellites still have some cold gas.  However, the LERG centrals are more efficient at converting the little cold gas they have to stars, while the cold gas of LERG satellites is relatively inhibited from condensing to form stars. 
\end{enumerate}
These results indicate that whether a radio galaxy is a satellite or central is correlated with the fraction of cold gas within the host galaxy. This is more relevant to the population of cold accretion-dominated HERGs, as satellites have a reduced amount of cold gas to be efficiently accreted onto the central black hole. Otherwise being a central or BCG plays no significant role in the evolution of radio galaxies.

On the environments of radio galaxies, we conclude the following:
\begin{enumerate}[i]
    \item Multiple LERGs can be hosted by a single dark matter halo, and they cluster similarly to the total galaxy population. Meanwhile the average number of HERGs hosted by the same dark matter halo is less than 1, and they are usually separated by $\ga$1~Mpc \edit{indicating that no 2 HERGs share a host dark matter halo}. Thus \simba\ predicts that the 1-halo clustering term for radio galaxies is dominated by LERGs.
    \item At low stellar masses ($M_*\la 10^{11}M_\odot$), HERGs and LERGs typically live in under-dense environments similar to that of star forming galaxies and comparable to non-radio galaxies at the same stellar masses. At high stellar masses, LERGs live in rich environments similar to non-radio and quiescent galaxies, reflecting the relations between black hole, stellar, and halo mass. The accretion of hot gas which creates LERGs is enabled by the hot circum-galactic medium in high density environments, while the HERGs with over-massive black holes typically reside in low density environments with cooler gas.
    \item For the radio luminosities probed in \simba, the richness of the environment of radio galaxies is independent of radio luminosity. This agrees with observations showing that the environments of radio galaxies are independent of radio luminosities at$\power\la 10^{24}\WHz$  \citep{Ching2017}.
    \item Radio galaxies live in more rich and dense environments than non-radio galaxies with accreting black holes at $\fedd \ga 10^{-6}$. There is, however, no significant dependence of the environment of either radio or non-radio galaxies on $\fedd$ for the radio luminosities probed by \simba.
    \item At $z=2$, radio galaxies reside in less rich and less dense environments than at $z=0$, and in environments similar to that of of non-radio galaxies at the same $\mstar$, $\power$, or $\fedd$. At these redshifts, high density environments are more able to support the cold gas accretion onto HERGs, presumably owing to more rapid cooling from denser gas.
\end{enumerate}

These results suggest that the observational relations currently seen between radio galaxies and their environments are biased owing to high radio luminosity populations. \simba\ makes novel and testable predictions for the evolution of low luminosity radio galaxies.  Next generation sensitive and deep continuum radio surveys such as MIGHTEE will shed light on the evolution of radio galaxies and the environments that fuel them~\citep{Whittam2022}, and provide more stringent constraints on the nature of radio galaxies in simulations such as \simba. 

\section*{Acknowledgements}
The authors would like to thank the anonymous reviewer for helpful comments, Emily Moravec for helpful discussions, and Philip Hopkins for making Gizmo public, and providing our group with early access. We thank Robert Thompson for developing Caesar, and the yt team for development and support of yt. We acknowldedge support from Newton Mobility Grant NMG-R1-180195 from the U.K. Royal Society. NT acknowledges support from the South African Radio Astronomy Observatory, which is a facility of the National Research Foundation, an agency of the Department of Science and Technology as well as support by the Medical Research Council [MR/T042842/1]. RD acknowledges support from Wolfson Research Merit Award WM160051 from the U.K. Royal Society. The \simba\
simulation was run on the DiRAC@Durham facility managed by
the Institute for Computational Cosmology on behalf of the STFC
DiRAC HPC Facility. The equipment was funded by BEIS capital funding via STFC capital grants ST/P002293/1, ST/R002371/1 and ST/S002502/1, Durham University and STFC operations grant ST/R000832/1. DiRAC is part of the National e-Infrastructure.

\section*{Data Availability}
The data used to derive the findings in this study are publicly available at \url{http://simba.roe.ac.uk}.



\bibliographystyle{mnras}
\bibliography{RadioCvS} 




\bsp	
\label{lastpage}
\end{document}